\documentclass[aps,preprint,showpacs,nofootinbib,amsmath]{revtex4}
\usepackage{graphicx}
\newcommand*{\tcaps}{\setlength{\baselineskip}{3ex}}
\newcommand*{\fs}[1]{#1\!\!\!/}
\usepackage{color}
\begin{document}

\title{
Coherent $\Theta^+$ and $\Lambda(1520)$ photoproduction off the
deuteron}

 \author{A.\,I.~Titov$^{a,b}$, B. K\"ampfer$^{a,c}$,
S. Dat\'e$^d$, and Y. Ohashi$^d$}

 \affiliation{
 $^a$Forschungzentrum Rossendorf, 01314 Dresden, Germany\\
 $^b$Bogoliubov Laboratory of Theoretical Physics, JINR,
  Dubna 141980, Russia\\
 $^c$ Institut f\"ur Theoretische Physik, TU~Dresden, 01062 Dresden, Germany\\
 $^d$Japan Synchrotron Radiation Research Institute, SPring-8,
 1-1-1 Kouto Mikazuki-cho Sayo-gun Hyogo 679-5198, Japan
 }

\begin{abstract}
We analyze an effect of the coherent $\Theta^+\Lambda(1520)$
photoproduction in $\gamma D$ interaction near the threshold. We
demonstrate that the coherence effect becomes manifest in a
comparison of the $nK^+$ invariant mass distribution when the
$pK^-$ invariant equals the $\Lambda(1520)$ mass. Our model
calculations indicate a sizeable contribution of resonant and
non-resonant background processes in the $\gamma D\to np K^+K^-$
reaction which generally exceed the contribution of the coherent
resonant channel. However, we find that the coherent
$\Theta^+\Lambda(1520)$ photoproduction is enhanced relative to
the background processes in the forward hemisphere of the $pK^-$
pair photoproduction. Moreover,  the coherence effect does not
depend on the $\Theta^+$ photoproduction amplitude and is defined
by the probabilities of the $\Lambda(1520)$ photoproduction and
the $\Theta^+\to NK$ transition. Therefore, this coherence effect
may be used as an independent method for studying the mechanism of
$\Theta^+$ production and $\Theta^+$ properties.
\end{abstract}


\maketitle

\section{Introduction}

 The first evidence for the  pentaquark hadron  $\Theta^+$ discovered by the
 LEPS collaboration at SPring-8~\cite{Nakano03} was subsequently
 confirmed in other experiments~\cite{OtherPenta}. However,
 some other experiments failed to find the $\Theta^+$ signal
 (for a review see~\cite{Hicks,Kabana05}). Since then  the situation
 concerning the existence of the  pentaquarks remained  controversial.
 Independent studies of the manifestation of a $\Theta^+$ state
 in different processes are, therefore, urgently desired.

 $\Theta^+$ photoproduction in the reaction $\gamma D\to np K^+K^-$
 seems to be very interesting and important~\cite{NakanoP04,TedeschiP04}.
 First, it allows to study simultaneously the $\gamma p\to\Lambda(1520)K^+$
 and $\gamma n\to \Theta^+K^-$ subreactions characterized by the similarity
 in the production mechanisms, i.e.
 both processes are described by the same set of
 the tree level Feynman diagrams~\cite{Hosaka0503,TEHN04,Hosaka0505}.
 Therefore, one hopes to define
 the ratio of $\Theta^+$ to $\Lambda(1520)$ photoproduction
 with minimal uncertainty of the
 production mechanisms, which is important for understanding
 the nature of $\Theta^+$.
 Second, in case of the $\gamma D$ interaction
 one can study qualitatively a new basic process -
 the coherent $\Theta^+\Lambda(1520)$ photoproduction.
 This reaction has its own physics interest and unambiguously
 will shed new light to pentaquark properties and
 the mechanism of the $\Theta^+$ photoproduction.

 It is commonly supposed now  that the total width of the $\Theta^+$ is
 as small as
 $\Gamma_{\Theta}\sim 1$~MeV~\cite{SmallWidth}, being much smaller
 than the total $\Lambda^*$ decay width,
 $\Gamma_{\Lambda^*}\simeq 15.6$~MeV~\cite{PDG} .
 (Throughout this paper, for simplicity, we use notation
 $\Lambda^*\equiv \Lambda(1520)$.)
 This means that the most promising way for
 studying the coherent $\Lambda^*\Theta^+$ production
 is to analyze the invariant $nK^+$ mass, $M_{nK^+}$, distribution at fixed
 invariant mass of the $pK^-$ pair, $M_{pK^-}$. The enhancement of the $\Theta^+$
 photoproduction, when $M_{pK^-}$ is in the
 vicinity of the $\Lambda^*$ mass, would indicate the manifestation
 of the coherent $\Lambda^*\Theta^+$ photoproduction.
 This particular channel will appear in strong competition
 with the resonant and non-resonant background processes.
 By the notation  "resonant process" we mean, for example, the $\Lambda^*$
 photoproduction from the proton inside the deuteron,
 when the neutron is a spectator, and
 similarly the $\Theta^+$ photoproduction from a neutron, when the
 deuteron's proton is a spectator.
 The notation "non-resonant" process denotes $K^+K^-$ photoproduction
 from a nucleon without excitation of  $\Lambda^*$ or $\Theta^+$.
 It is clear that the coherent photoproduction and the background
 processes must be analyzed together using the same theoretical
 approaches. This allows to define the kinematical conditions
 where the coherent channel manifests itself clearly above
 strong background processes.

 The aim of the present paper is to discuss these important aspects.
 Our model includes the elementary subprocesses of $\gamma N\to
 \Lambda^*K$ and $\gamma N\to \Theta^+\bar K$ reactions. For the latter
 ones
 we  use a model based on the effective Lagrangian approach
 of Ref.~\cite{TEHN04} which is, generally speaking, similar to the
 models developed by other authors in
 Refs.~\cite{Hosaka03,OKL031,NT03,
 LiuKo031,Zhao03,ZhaoKhal04,Oh-2,
 CloseZhao,Roberts04,Mart:2004at,Oh:2004wp}.
 All these approaches predict the approximate equality of the
 cross sections of the $\gamma n\to \Theta^+ K^-$ and
 $\gamma p\to \Theta^+ \bar K^0$ reactions. This
 equality may be changed into a  suppression of the
 $\gamma p\to \Theta^+ \bar K^0$ transition~\cite{Hosaka0503,Vita}.
 However, we are going to demonstrate that the amplitude of the
 coherent $\Lambda^*\Theta^+$ photoproduction,
 when $\Lambda^*$ is produced in the forward
 hemisphere in the $\gamma D$ center of mass system, is
 defined by the product of the $\Lambda^*$ photoproduction
 amplitude in $\gamma N$ interaction and the amplitude of the
 $\Theta^+\to NK$ transition.  In other words, the coherence effect
 of the $\Lambda^*\Theta^+$ photoproduction in the forward hemisphere
 does not depend on the $\Theta^+$ photoproduction amplitude and
 remains finite even if the cross section of the
 $\gamma p\to \Theta^+ \bar K^0$ reaction is vanishing.
 The coherence effect in the backward hemisphere is sensitive
 to the $\Theta^+$ photoproduction amplitude, and it is
 suppressed in parallel with the suppression of the
 $\gamma p\to \Theta^+ \bar K^0$ reaction.

 Our paper is organized as follows.
 In Sec.~II, we discuss the resonant $\Theta^+$ and $\Lambda^*$
 photoproduction from a nucleon.
 In Sec.~III, we consider the coherent $\gamma D\to \Lambda^*\Theta^+$
 reaction. Our model is similar to the approach of
 Ref.~\cite{gammaD}, developed for coherent $\Theta^+\Lambda(\Sigma^0)$
 photoproduction from a deuteron.
 In Sec.~IV, we discuss
 the  background processes. We start thereby from an analysis of
 the non-resonant background in "elementary" $\gamma N\to\Theta^+\bar K$
 and $\gamma N\to\Lambda^* K$ reactions. Then we apply these
 subprocesses to an analysis of the background spectator channels.
 Finally, we estimate the contribution of the coherent semi-resonant
 processes, which differ from the coherent photoproduction by the
 replacement of one hyperon by  $NK$ or $N\bar K$  pairs.
 The results of our numerical calculations are presented in Sec.~V.
 The summary is given in Sec.~VI. In
 Appendix~A, we show an explicit form of the transition operators
 for the resonance amplitude.

\section{photoproduction from a nucleon}

\subsection{\boldmath{$\Theta^+$} photoproduction}

\begin{figure}[h!]
{\centering
 \includegraphics[width=.42\textwidth]{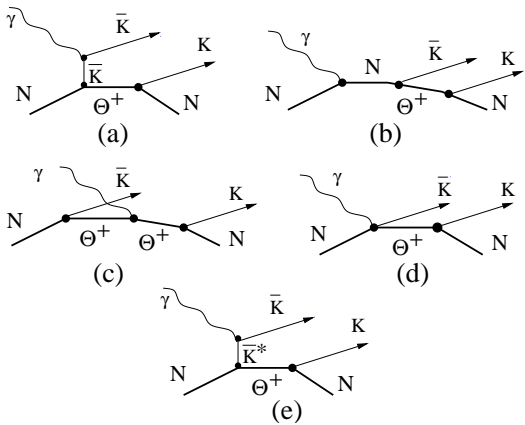}
 \caption{\label{fig:1}\tcaps%
 Tree level diagrams for the reaction
 $\gamma N\to \Theta^+\bar{K}\to NK\bar{K}$.}}
\end{figure}

 The main diagrams for the amplitude of the resonance $\Theta^+$ photoproduction
 in the reaction $\gamma N\to NK\bar K$ are shown in Fig.~\ref{fig:1}.
 We neglect here the contribution resulting from the photon
 interacting with the final decay vertex~\cite{NT03}. In view of the chosen
 kinematics, where the invariant mass of the final $KN$ pair is
 near the resonance position, this is a good approximation since in the
 neglected graphs the $\Theta^+$ is far off-shell and the graphs of
 Figs.~\ref{fig:1}~a - d dominate the resonance contribution. From a
 formal point of view gauge invariance is lost without contributions
 arising from the electromagnetic interaction in the decay
 vertex.
 However, following Ref.~\cite{hhgauge} for the initial
 photoproduction process, we construct an overall conserved
 current by an appropriate choice of the contact term of
 Fig.~\ref{fig:1}d.

 In this section
 $k$, $p$, $q$, $\bar q$, and $p'$ denote the four-momenta of the incoming photon,
 the initial nucleon, the outgoing $K$ and $\bar{K}$ mesons, and
 the recoil nucleon, respectively. The standard Mandelstam
 variables for the virtual $\Theta^+$ photoproduction are defined
 by $t= (\bar{q}-k)^2$, $s \equiv W^2 = (p+k)^2$.  The $\bar{K}$
 meson production angle $\theta$ in the center-of-mass system (c.m.s.)
 is given by $\cos\theta = \mathbf{k} \cdot \mathbf{\bar{q}} /(
 |\mathbf{k}| |\mathbf{\bar{q}}|)$, and the corresponding solid angle is
 $\Omega$. We consider the integrated
 $\Theta^+$ decay distribution.
 The differential cross section
 $\gamma N\to \Theta^+\bar K\to NK\bar K$ as a function
 the $\bar K$ meson production angle and
 $NK$ invariant mass, $M_{nK^+}$,
 at the resonance position with $M_{nK^+}=M_\Theta=1.54$~GeV
 is related to the cross section of the $\Theta^+$ photoproduction in the
 $\gamma N\to \Theta^+\bar{K}$ reaction as
\begin{eqnarray}
 \frac{d\sigma^R_{fi}}{d \Omega\, dM_{nK^+}}\Bigg|_{M_{nK^+}=M_\Theta}=
 \frac{1}{\pi\Gamma_{\Theta}}\frac{d\sigma^{\Theta^+}_{fi}}{d
 \Omega}
 \label{csR}
\end{eqnarray}
with $\Gamma_\Theta$ as $\Theta^+$ decay width and
\begin{eqnarray}
 \frac{d\sigma^{\Theta^+}_{fi}}{d\Omega}=
 \frac{1}{64\pi^2s}\,\frac{p_{\rm out}}{p_{\rm in}}\,
 \frac14\sum_{m_i,m_f,\lambda_\gamma}
 | A^{\Theta^+}_{m_f;m_i,\lambda_\gamma}|^2~.
 \label{csT}
\end{eqnarray}
  Here, $A^{\Theta^+}$ is the $\Theta^+$ photoproduction amplitude
  in the $\gamma N\to \Theta^+\bar{K}$ reaction,
  $m_i$ and $m_f$ are the nucleon and $\Theta^+$ spin projections, respectively,
  and $\lambda_\gamma$ denotes the incoming photon helicity; $p_{\rm in}$ and $p_{\rm out}$
  are the relative momenta in the initial and the final states in c.m.s.,
  respectively. Further on we
  will concentrate on the calculation of $A^{\Theta^+}$.
  For simplicity, in this analysis we limit our
  consideration to the {isoscalar}, spin-1/2 $\Theta^+$.
  Generalization for higher spin~\cite{Hosaka0505}
  may be done in a straightforward manner.

  The effective Lagrangians which define the Born terms
  for the diagrams   shown in Fig.~\ref{fig:1}a - d
  are discussed in many papers (for references see Ref.~\cite{TEHN04}).
  Note that different phase conventions
  are often employed.  Therefore, for the sake of definiteness,
  we list here the effective Lagrangians used in the present
  work\footnote{%
Throughout this paper, isospin operators will be suppressed in all
Lagrangians and matrix elements for simplicity. They can be easily
accounted for in the corresponding coupling constants.}:
\begin{subequations}
\label{Leff}
\begin{eqnarray}
 {\cal L}_{\gamma KK}^{} &=& i e\,
 (K^- \partial^\mu K^+
 - K^+\partial^\mu K^-)A_\mu~,
 \label{AKK}\\
{\cal L}^{}_{\gamma\Theta\Theta}
  &=& - e \,\bar \Theta \left(\gamma_\mu
  -\frac{\kappa_\Theta}{2M_\Theta}\sigma_{\mu\nu}\partial^\nu\right) A^\mu\Theta~,
  \label{gTT}\\
{\cal L}^{}_{\gamma NN}
  &=& - e \,\bar N \left(e_N\gamma_\mu -
  \frac{\kappa_N}{2M_N}\sigma_{\mu\nu}\partial^\nu\right) A^\mu N~,
  \label{gNN}\\
 {\cal L}^{\pm[\text{pv}]}_{\Theta NK}
  &=& \mp \frac{g_{\Theta NK}}{M_\Theta \pm M_N}
  \bar \Theta \Gamma^{\pm}_\mu (\partial^\mu K) N  + \text{h.c.}~,
  \label{TNK}\\
 {\cal L}^{[\rm pv]}_{\gamma\Theta NK}
  &=& -i \frac{eg_{\Theta NK}}{M_\Theta \pm M_N}
  \bar \Theta \Gamma^{\pm}_\mu A^\mu K N + \text{h.c.}~,
  \label{gTNK}\\
  {\cal L}^{\pm[\text{ps}]}_{\Theta NK}
  &=& -i {g_{\Theta NK}}
  \bar \Theta \Gamma^{\pm}\,K\, N  + \text{h.c.}~,
  \label{TNKps}\\
  {\cal L}_{\gamma K K^*}^{} &=& \frac{e g_{\gamma KK^*}}{M_{K^*}}
  \epsilon^{\alpha\beta\mu\nu}
  \partial_\alpha A_\beta
  \partial_\mu \bar{K}^{*}_\nu K  +  \text{h.c.}~,
  \label{gammaKK*}\\
  {\cal L}_{\Theta NK^{*}}^{\pm} &=& - g_{\Theta NK^*}
  \,\bar \Theta \,\Gamma^{\mp}\left(\gamma_\mu  -
  \frac{\kappa^*}
  {M_\Theta+M_N}\sigma_{\mu\nu}\partial^\nu \right) \bar{K}^{*\mu} N
  + \text{h.c.}~,
  \label{TNK*}
  \end{eqnarray}
  \end{subequations}
 where $A^\mu,\,\Theta$, $K$, and $N$ are the photon, $\Theta^+$,
 kaon, and the nucleon fields, respectively,
 $K^*$ stands for the vector kaon field;
 $\Gamma^\pm_\mu\equiv\Gamma^{\pm} \gamma_\mu$
 (with $\Gamma^+=\gamma_5$ and $\Gamma^-=1$ for positive and
 negative parity, respectively),
  $e_p=1$, $e_n=0$,  and $\kappa_{N}$ denotes the
 nucleon anomalous magnetic moment ($\kappa_p=1.79$ and
 $\kappa_n=-1.91$), $\kappa_\Theta$ stands
 for the anomalous magnetic moment of $\Theta^+$
 and $\kappa^*$ denotes the tensor coupling of nucleon and strange
 vector mesons.
  The superscripts "PS" and "PV" correspond to
 the pseudo-scalar and pseudo-vector $\Theta^+NK$ coupling schemes.
 Equation~(\ref{gTNK}) describes the contact (Kroll-Ruderman)
 interaction in the pseudo-vector coupling scheme (see
 Fig.~\ref{fig:1}d), which does not appear in case of the
 pseudo-scalar coupling (cf. Eq.~(\ref{TNKps})).

In calculating the invariant amplitudes we dress the vertices by
form factors. In the present tree-level approach and within our
chosen kinematics, only the lines connecting the electromagnetic
vertex with the initial $\Theta^+KN$ vertex correspond to
off-shell hadrons. We describe the product of both the
electromagnetic and the hadronic form-factor contributions along
these off-shell lines by the covariant phenomenological function
\begin{eqnarray}
 F(M,p^2)=\frac{\Lambda^4}{\Lambda^4 + (p^2-M^2)^2}~,
 \label{FF}
\end{eqnarray}
where $p$ is the corresponding off-shell four-momentum of the
virtual particle, $M$ denotes its mass, and $\Lambda$ stands for
the cut-off parameter. The electromagnetic current of the complete
amplitude is conserved by making the initial photoproduction
process gauge invariant. To this end, we apply the gauge
invariance prescription by Haberzettl~\cite{hhgauge} with the
modification by Davidson and Workman~\cite{DavWork} to construct a
contact term for the initial process $\gamma N \to \Theta^+
\bar{K}$ free of kinematical singularities. We emphasize that
contributions of the latter type are necessary even for pure
pseudo-scalar coupling.

 Since the coupling scheme and the $\Theta^+$ parity are unknown
 one has to define the corresponding parameters in such a way
 to get the corresponding cross sections
 independently of $\Theta^+$ parity
 and coupling scheme. We follow Ref.~\cite{TEHN04}, where
 parameters of the model are fixed by a comparison of the
 resonant $\Theta^+$ photoproduction cross section
 and non-resonant background with experiment and it is shown that
 one can find such a parameter set which parallels the prediction
 for PS and PV couplings and for positive and negative
 $\Theta^+$ parity states as well,
 at least for the unpolarized, single and double
 polarization spin observables.
 Therefore, we can limit the present analysis
 to the PS coupling and a positive $\Theta^+$ parity.

The resonance amplitudes obtained for the $\gamma n$ and $\gamma
p$ reactions read
\begin{subequations}
\label{res_ampl}
\begin{eqnarray}
 &&  A^{\Theta^+}_{fi}(\gamma n)=\bar u_{\Theta}(p_\Theta)
 \left[
  {{{\cal M}^s}}_\mu +
  {{{\cal M}^t}}_\mu +
  {{{\cal M}^u}}_\mu +
  {{{\cal M}^c}}_\mu +
  {{{\cal M}^t}}_\mu(K^*)
  \right]
 u_n(p)\,\varepsilon^\mu,\\
&&  A^{\Theta^+}_{fi}(\gamma p)=\bar u_{\Theta}(p_\Theta)
  \left[
 {{{\cal M}^s}}_\mu +
  {{{\cal M}^u}}_\mu +
  {{{\cal M}^c}}_\mu +
  {{{\cal M}^t}}_\mu(K^*)
  \right]
 u_p(p)\,\varepsilon^\mu~.
\end{eqnarray}
\end{subequations}
The explicit forms of the transition operators  ${\cal M}^{i}_\mu$
for the $\gamma n\to \Theta^+ K^-$ and $\gamma p\to
\Theta^+\bar{K}^0$ reactions are exhibited in Appendix A.
%

 For a positive $\Theta^+$ parity
 the coupling constant $g_{\Theta NK}$
 is found from the $\Theta^+$ decay width as
\begin{eqnarray}
\Gamma_{\Theta}=\frac{[g_{\Theta NK}]^2 p_F}{2\pi M_\Theta}
(\sqrt{M_N^2 + p_F^2}- M_N)~.
\end{eqnarray}
 We choose a small width, $\Gamma_\Theta=1$~MeV~\cite{SmallWidth}, assuming
 that the observed width in the invariant mass distribution is
 determined by the experimental resolution.
 The magnitude of the coupling constant $g_{\gamma KK^*}$ is
 extracted from the width of the $K^*\to \gamma K$
 decay~\cite{PDG}. Its sign is fixed by SU(3) symmetry. This
 delivers
 $eg_{\gamma K^{0}K^{*0}}=-0.35$ and $eg_{\gamma K^{\pm}K^{*\pm}}=
 0.23$. The contribution of the $s$-channel (Fig.~\ref{fig:1}b)
 is  small causing to a
 rather weak dependence of the total amplitude on the tensor
 coupling $\kappa_\Theta$ in the $\gamma \Theta\Theta$ vertex
 within a ``reasonable" range of $0\lesssim|\kappa_\Theta|
 \lesssim0.5$~\cite{MagMom}. Therefore, we can choose
 $\kappa_\Theta=0$.
  The coupling constant $g_{\Theta NK^*}$ is written as
  $g_{\Theta NK^*}=\alpha_\Theta g_{\Theta NK}$, where the parameter $\alpha_\Theta$
  depends on the choice of the tensor coupling $\kappa^*$ in
  Eq.~(\ref{TNK*}) and cut-off parameters $\Lambda_{K^*}$ in the form factors
  of the $K^*$ exchange amplitude.  Increasing
  value of $\Lambda_{K^*}$ leads to a decreasing $\alpha_\Theta$.
  Following Ref.~\cite{TEHN04}
  we use $\Lambda_{K^*}=1.5$~GeV and $\alpha_\Theta=1.875$ at $\kappa^*=0$.
  This value of $\alpha_\Theta$ is close to the quark model estimates
  $\alpha_\Theta=\sqrt{3}$~\cite{QMKK*}.

  Another cut-off parameter, $\Lambda_B$, defines the Born terms of
  the $s$-, $u$-, and $t$-channels and the current-conserving
  contact terms. Note that the inclusion of the $\Sigma$ and $\Lambda$
  photoproduction processes~\cite{LambdaSigma} results in a larger
  ambiguity in the choice of $\Lambda_B$ which varies from 0.5 to 2
  GeV depending on the coupling scheme and the method of conserving the
  electromagnetic current etc.
  The analysis of the vector meson
  photoproduction~\cite{TitovLee} and $\gamma n\to \Theta^+ K^-$ favor
   a small value of the cut-off, $\Lambda_B\simeq 0.5$ GeV. For
  the $\gamma p\to \Theta^+ \bar K^0$ reaction
  the $K^*$ exchange channel remains to be dominant at $\Lambda_B\leq1.5$ GeV
  and, therefore, in this paper we use a "universal" value, $\Lambda_B\simeq 0.5$~GeV,
  for all Born terms.

\begin{figure}[h!]
{\centering
 \includegraphics[width=.4\textwidth]{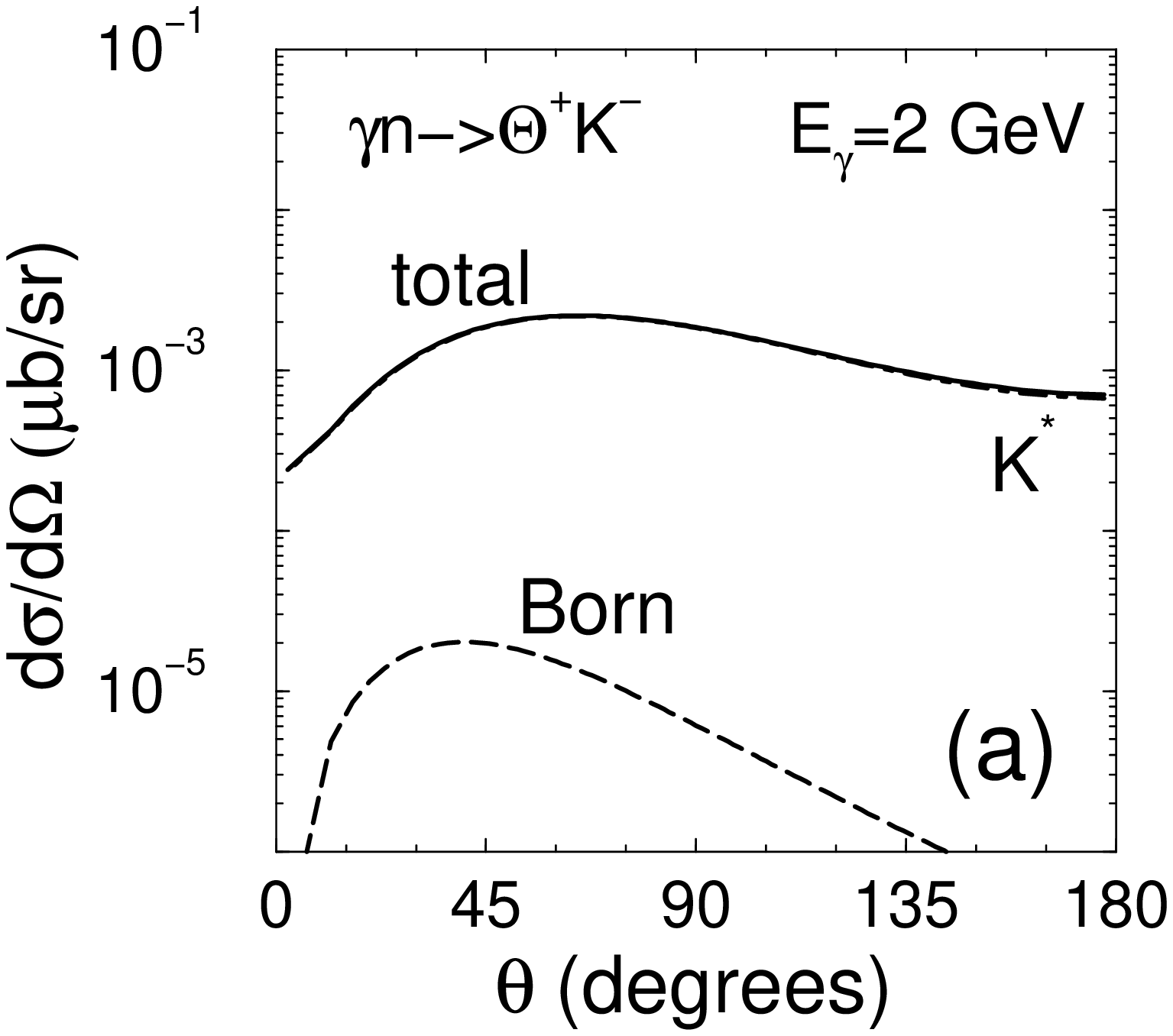}\qquad
 \includegraphics[width=.4\textwidth]{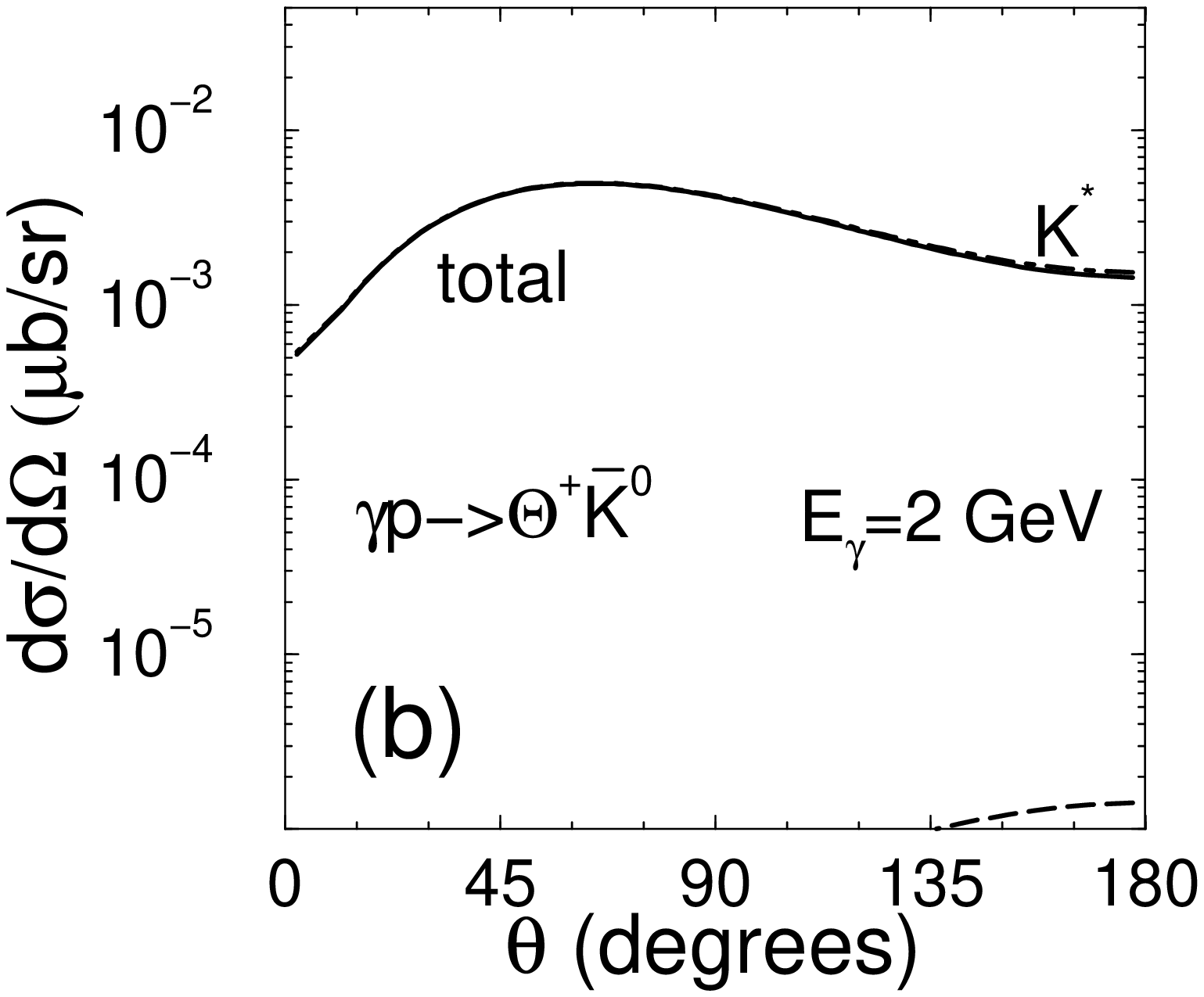}
 \caption{\label{fig:2}\tcaps%
 The differential cross section of the reaction $\gamma n\to \Theta^+K^-$ (a) and
 $\gamma p\to \Theta^+\bar K^0$ (b) at $E_\gamma=2$  GeV. The notation
 "Born" correspond to the coherent sum of the $t,s,u$-exchange diagrams
 and the contact term, shown in
 Fig.~\protect\ref{fig:1}a - d, respectively.
 Solid curves indicate the total of all contributions. The
contribution of K* exchange is indicated and shown as dash-dotted
curves which almost overlap with solid curves.}}
 \end{figure}

 In Fig.~\ref{fig:2} we exhibit the differential cross sections of the
 reactions $\gamma n\to \Theta^+K^-$ (a) and
 $\gamma p\to \Theta^+\bar K^0$ (b) in the c.m.s. at $E_\gamma=2$  GeV. One can see
 that the $t$-channel $K^*$ exchange depicted in
 Fig.~{\ref{fig:1}e
 gives the dominant contribution compared to the Born terms
 shown in Fig.~\ref{fig:1}a - d in both reactions.

\subsection{\boldmath{$\Lambda(1520)$} photoproduction}

 The main diagrams
 for the amplitudes of the excitation of the $\Lambda$ hyperon
 in the $\gamma N\to NK\bar K$ reaction at low energies
 are shown in Fig.~\ref{fig:3}. Similarly to the $\Theta^+$
 photoproduction we neglect the photon interaction within the decay
 vertex and restore the gauge invariance by a proper choice of the
 contact terms.
 The  Mandelstam variables for the virtual $\Lambda^*$ photoproduction are defined
 by $t= ({q}-k)^2$, $s \equiv W^2 = (p+k)^2$.  The ${K}$
 meson production angle $\theta$ (in $\gamma p$ c.m.s.)\
 is given by $\cos\theta = \mathbf{k} \cdot \mathbf{{q}}
 /(|\mathbf{k}| |\mathbf{{q}}|)$.
 \begin{figure}[h!]
{\centering
 \includegraphics[width=.42\textwidth]{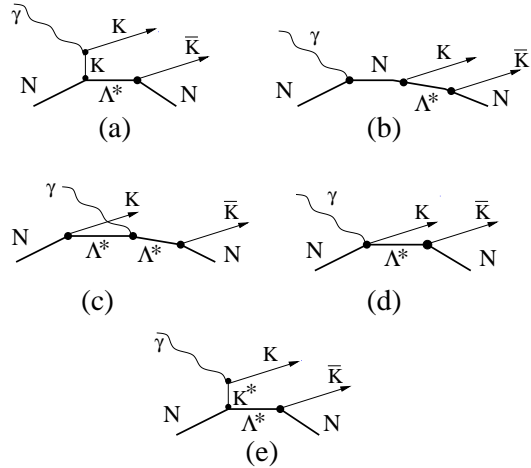}
 \caption{\label{fig:3}\tcaps%
 Tree level diagrams for the reaction
 $\gamma N\to \Lambda^*{K}\to NK\bar{K}$.}}
\end{figure}

 For the description of the $\Lambda^*$ excitation with $J^P={\frac{3}{2}}^-$
 we use the following effective Lagrangians~\cite{3-2R,TitovLee}
 \begin{subequations}
\label{LLeff}
\begin{eqnarray}
 {\cal L}^{}_{\Lambda^* NK}
  &=& \frac{g_{\Lambda^* NK}}{M_{\Lambda^*}}
  \,{{\bar\Lambda^*}}_\mu\,\theta^{\mu\nu}(Z)\,(\partial_\nu\bar K)\,\gamma_5
  N  \,+ \,\text{h.c.}~,
  \label{LNK}\\
 {\cal L}^{}_{\gamma\Lambda^* NK}
  &=& -i \frac{eg_{\Lambda^* NK}}{M_\Lambda^*}
  \,{{\bar {\Lambda}^*}}_\mu\, \gamma_5 A^\mu \bar K N \,+ \,\text{h.c.}~,
  \label{gLNK}\\
 {\cal L}_{\Lambda^* NK^{*}}^{\pm} &=& i \frac{g_{\Lambda^*
 NK^*}}{M_{\Lambda^*}}
  \,{{\bar {\Lambda}^*}}_\mu\,\theta^{\mu\nu}(Y)\gamma^\lambda {F_{\bar K}}_{\lambda\nu} N
\,+\,  \text{h.c.}~,
 \label{LNK*}
  \end{eqnarray}
  \end{subequations}
 where $\Lambda^*$ is the $\Lambda$(1520) field,
 $M_{\Lambda^*}$ denotes the $\Lambda^*$ mass,
 $F_K^{\mu\nu}$ is related to the
 vector $K^*$ meson field as $F_K^{\mu\nu}=\partial^\nu {K^*}^\mu
 - \partial^\mu {K^*}^\nu$. The operator $\theta_{\mu\nu}(X)$
 is a function of the "off-shell" parameter $X$:
 $\theta_{\mu\nu}(X)=g_{\mu\nu}-(\frac{1}{2}+X)\gamma_\mu\gamma_\nu$.
 In this paper we consider such a kinematics where the invariant mass of the outgoing
 $N\bar K$ pair is close to $M_{\Lambda^*}$, $\Lambda^*$ is almost
 on-shell, and therefore,  the contribution from terms proportional
 to $\gamma_\mu\gamma_\nu$ in $\theta_{\mu\nu}(X)$
 disappears. This means that $\theta_{\mu\nu}(X)$ may be replaced
 by $g_{\mu\nu}$.
 We assume a vanishing value of the anomalous magnetic moment of $\Lambda^*$
 and, therefore, neglect the $\Lambda^*\gamma$ interaction, and,
 correspondingly,
 the contribution of the $u$-channel shown in Fig.~\ref{fig:3}c.
 All vertices are dressed by the form factors similarly to the
 case of the $\Theta^+$ photoproduction with the same cut-off parameters.
 The amplitudes for the $\gamma p\to\Lambda^*K^+$
 and $\gamma n\to\Lambda^*K^0$ reactions read
\begin{subequations}
\label{res_amplL}
\begin{eqnarray}
 &&  {A^{\Lambda^*}}_{fi}(\gamma p)=\bar u_{\Lambda^*}^\sigma(p_\Lambda^*)
 \left[
  {{{\cal M}^s}}_{\sigma\mu} +
  {{{\cal M}^t}}_{\sigma\mu} +
  {{{\cal M}^c}}_{\sigma\mu} +
  {{{\cal M}^t}}_{\sigma\mu}(K^*)
  \right]
 u_p(p)\,\varepsilon^\mu,\\
 && {A^{\Lambda^*}}_{fi}(\gamma n)=\bar u_{\Lambda^*}^\sigma(p_\Lambda^*)
  \left[
 {{{\cal M}^s}}_{\sigma\mu} +
  {{{\cal M}^t}}_{\sigma\mu}(K^*)
  \right]
 u_n(p)\, \varepsilon^\mu~.
\end{eqnarray}
\end{subequations}
 The explicit transition operators
 ${\cal M}^{i}_{\sigma\mu}$ for these reactions are listed in Appendix A.

 The coupling constant $g_{\Lambda^* NK}$  is found from
the $\Lambda^*$ decay width,
\begin{eqnarray}
\Gamma_{\Lambda^*\to N\bar K}=\frac{[g_{\Lambda^* NK}]^2 p^3_F
}{6\pi M_{\Lambda^*}^3} (\sqrt{M_N^2 + p_F^2}- M_N)~,
\end{eqnarray}
where $p_F$ is $\Lambda^*\to N\bar K$-decay momentum. Taking
$\Gamma_{\Lambda^*\to N\bar K}\simeq 0.45\times
15.6$~MeV~\cite{PDG}, one finds $|g_{\Lambda^* NK}|=32.6$.

 Analog to the above considered $\Theta^+$ photoproduction we denote
 $g_{\Lambda^* NK^*}=\alpha_{\Lambda^*}g_{\Lambda^* NK}$.
 The  parameter $\alpha_{\Lambda^*}$ must be defined by a comparison
 of calculated cross sections
 with experimental data at $E_\gamma\sim  2$ GeV. However, the available
 experimental data for the $\gamma p\to \Lambda^*K^+$ reaction cover
 the energy range $E_{\gamma}=2.8-4.8$~(GeV)~\cite{Barber1980},
  beyond the applicability of the effective Lagrangian
 formalism.  Thus, in this region the total cross section decreases
 with energy as $E_\gamma^{-2.1}$, whereas the amplitudes of
 Eq.~(\ref{res_amplL}) predict a strong increase.
 The energy dependence at high energy is reasonably well
 described by the Regge phenomenology. Since
 the $\Lambda^*$ decay angular distribution
 supports the dominance of
 the $t$-channel natural parity exchange processes,
 one can assume that the dominant contribution to the
 $\Lambda^*$ photoproduction at high energy comes from
 the leading $K^*$ trajectory~\cite{Collins}.
 The corresponding amplitude is obtained from the $t$-channel
 $K^*$ meson exchange in Eq.~(\ref{res_amplL}) by the Reggezation
 of the $K^*$ meson exchange propagator, i.e.
\begin{eqnarray}
 \frac{1}{t-M_{K^*}^2}\to
 \gamma(t)\,\left(\frac{s}{s_0}\right)^{\alpha(t)}~,
 \label{R1}
\end{eqnarray}
where $\alpha(t)=\alpha(0) +\alpha'\,t$ is the Regge trajectory
and $\gamma(t)$ denotes the normalization function
\begin{eqnarray}
 &&\gamma(t)=C_R({\rm Tr}[R\,R^\dagger])^{-1}~,\nonumber\\
 &&R=\bar u_{\Lambda^*}^\sigma(p_\Lambda^*)
  \left[\varepsilon^{\nu\mu\alpha\beta}\,k^\nu {q'}^\alpha
  ({q'}_\sigma \gamma^\beta - \fs{q'}g_{\sigma}^{\beta})\
  \right]
 u_n(p)\,\varepsilon^\mu
 \label{R2}
\end{eqnarray}
 with $q'=p_{\Lambda^*}-p$.
 In the following we assume that at  energies near the threshold, the
 production amplitude is defined by the effective Lagrangian
 model of Eq.~(\ref{res_amplL}), $A^{\Lambda^*}_{eff.\,L.}$, whereas
 at high energies it is described by the Regge phenomenology,
 $A^{\Lambda^*}_R$, as
\begin{eqnarray}
 A^{\Lambda^*}=A^{\Lambda^*}_{eff.\,L.}\,\theta(E_0-E_\gamma) +
 A^{\Lambda^*}_R\,\theta(E_\gamma-E_0)~.
 \label{R3}
\end{eqnarray}
 We take $E_0=2.3$ GeV as matching point between the two regimes.
 The choice of parameters in Eq.~(\ref{R2}) as
 $s_0=1$ GeV, $\alpha(t)=-0.1 +0.9t$ and $C_R=29.6$
 gives a satisfactory description of the high energy data,
 as exhibited in Fig.~\ref{fig:4} for the differential cross section
 at $E_\gamma=3.7$~GeV.
 \begin{figure}[h!]
{\centering
 \includegraphics[width=.4\textwidth]{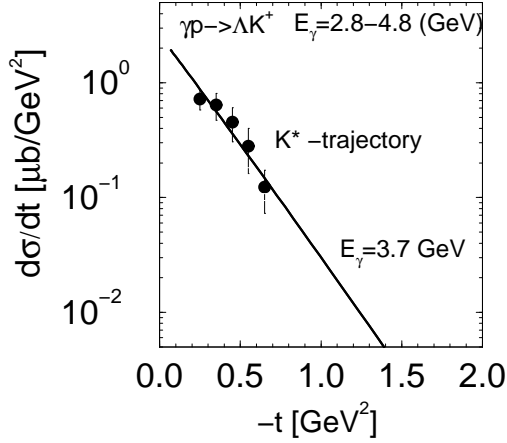}
 \caption{\label{fig:4}\tcaps%
 Differential cross section of the reaction $\gamma p\to \Lambda^+K^+$
 at $E_\gamma=3.7$~GeV. Experimental data from Ref.~\protect\cite{Barber1980}. }}
 \end{figure}

 In Fig.~\ref{fig:5}
 we show the energy dependence of the total cross section.
\begin{figure}[h!]
 {\centering
 \includegraphics[width=.4\textwidth]{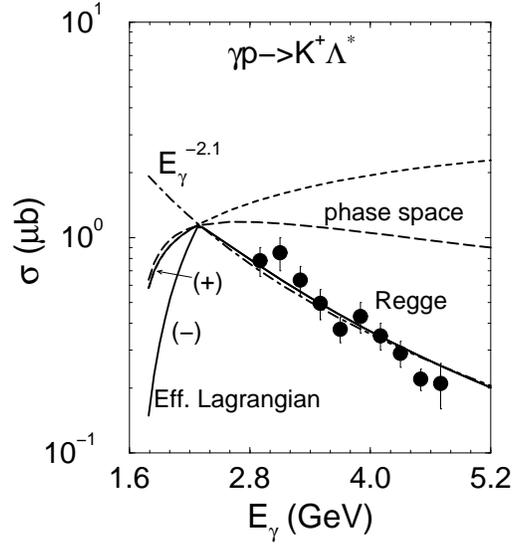}
 \caption{\label{fig:5}\tcaps%
 The total cross section of the reaction $\gamma p\to \Lambda^*K^+$
 as a function of the photon energy. The experimental data
 are taken from Ref.~\protect\cite{Barber1980}.
 The dot-dashed curve is the fit of this data
 $\sigma\simeq6.55\,E_\gamma^{-2.1}$ ($\mu$b). The long dashed curve
 represents the cross section when the amplitude is taken to be
 constant. The solid curves corresponds to the amplitude of
 Eq.~(\ref{R3}). The signs "$\pm$" corresponds to the sign
 of $\alpha_{\Lambda^*}$. The dashed curve describes
 the extrapolation of the effective Lagrangian model to the high
 energy region.}}
 \end{figure}
 The dot-dashed curve is the fit of the data
 $\sigma\simeq 6.55\,(E_\gamma/{\rm GeV})^{-2.1}$ ($\mu$b) from~\cite{Barber1980}.
 For illustration we also show the cross section calculated
 with a constant amplitude where the energy dependence
 is defined by the phase space volume alone.
 The strength parameter $\alpha_{\Lambda^*}$ is adjusted
 by fitting the calculated cross section
 to the experimental extrapolation (dot-dashed curve)
 at the normalization point. Two solutions
 $\alpha_{\Lambda^*}=+0.372$ and $-0.657$ result in
 two different energy dependencies of the cross section
 at low energy. Both solutions exceed the experimental
 data above the normalization point.
 The solution with positive $\alpha_{\Lambda^*}$ at low energies is
 close to the pure phase space dependence shown by the long-dashed
 curve.
\begin{figure}[h!]
{\centering
 \includegraphics[width=.4\textwidth]{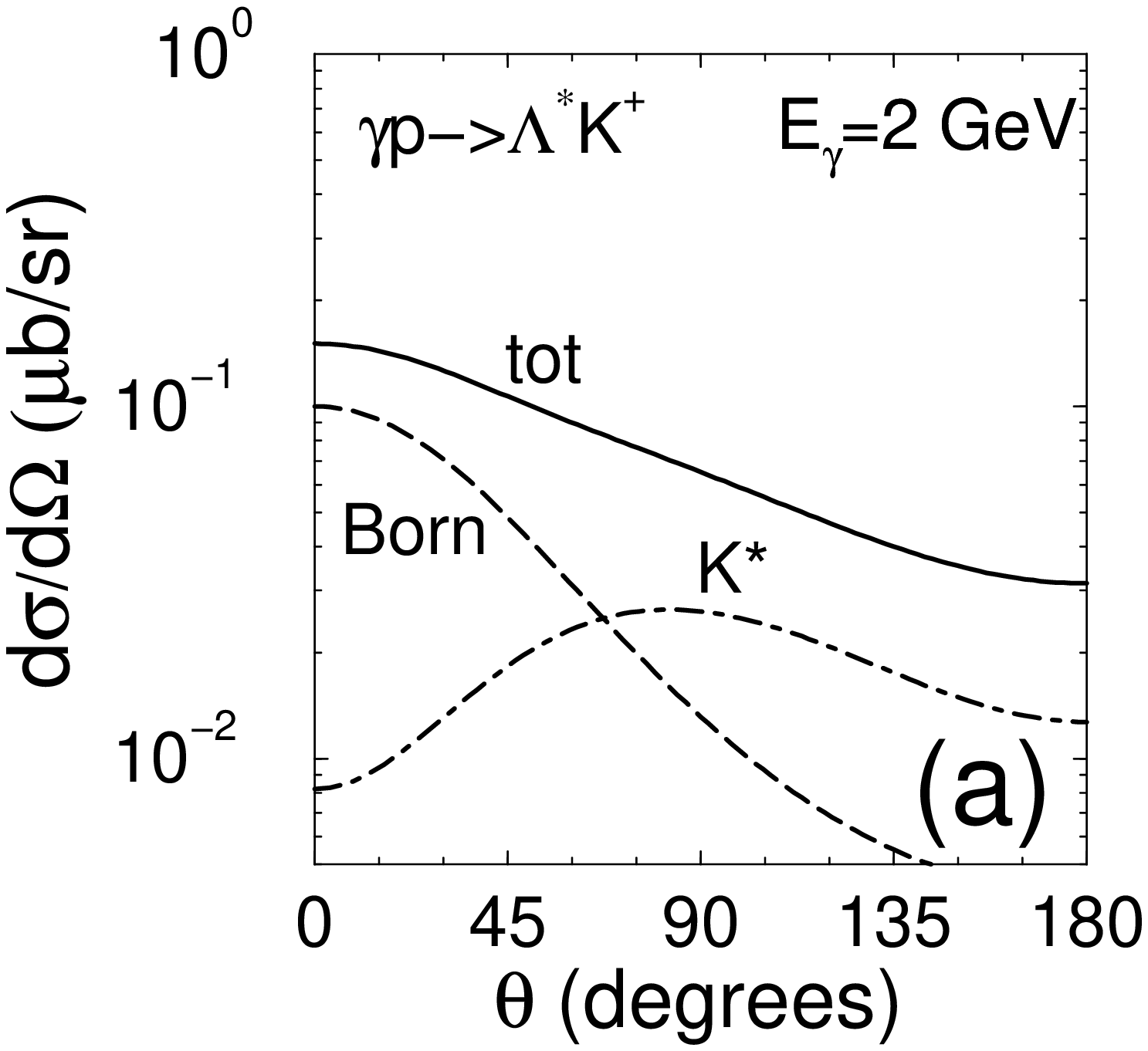}\qquad
 \includegraphics[width=.4\textwidth]{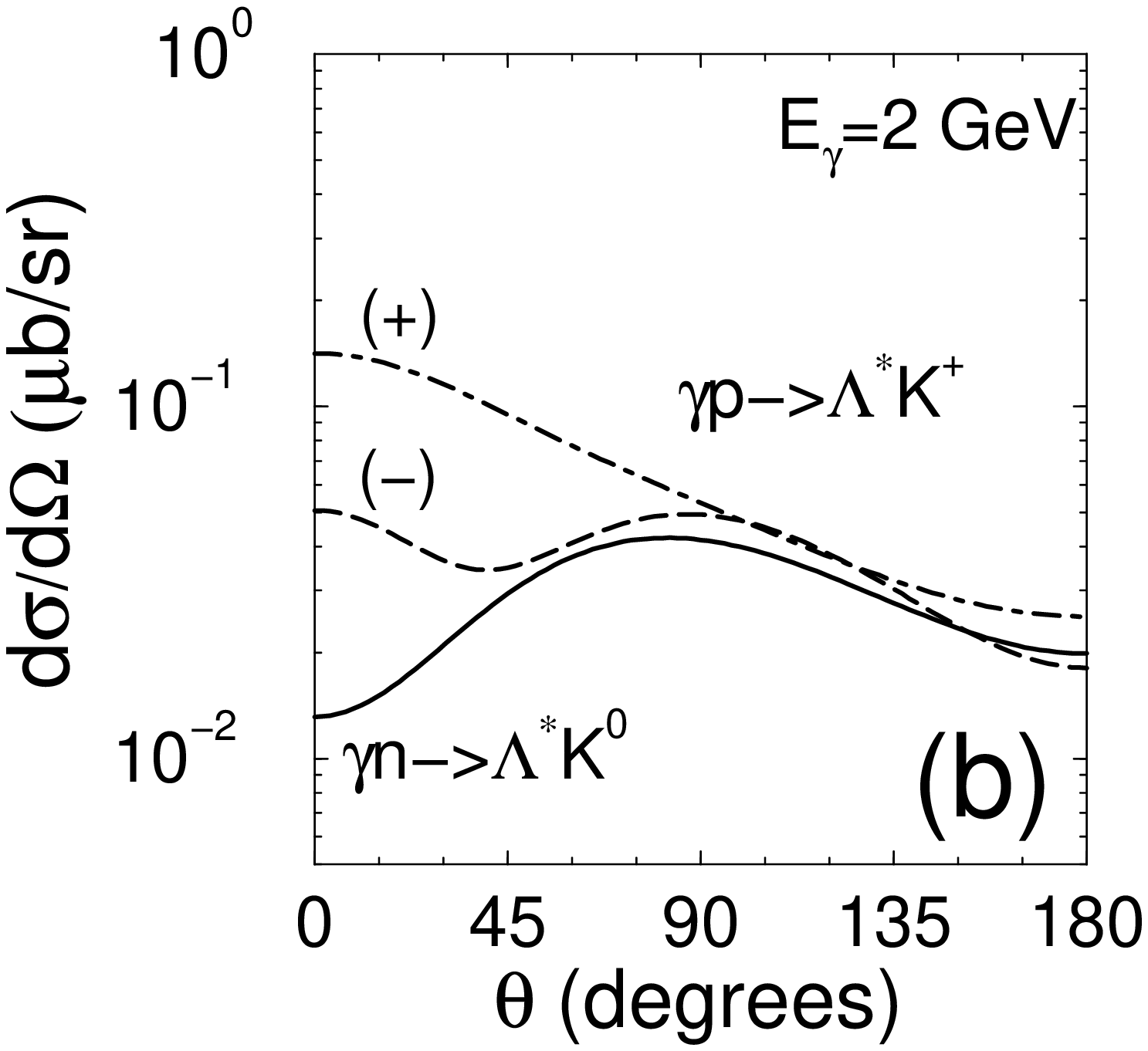}
 \caption{\label{fig:6}\tcaps%
 (a) The differential cross section of the $\gamma p\to \Lambda^*K^+$
  reaction at $E_\gamma=2$~GeV. The notation
 "Born" corresponds to the coherent sum of the $t,s$-exchange
 diagram
 and the contact term, shown in
 Fig.~\protect\ref{fig:3}a, b, and d, respectively.
 (b) The differential cross section of the $\gamma n\to \Lambda^*K^0$
  reaction (solid curve) and $\gamma p\to \Lambda^*K^+$ reaction
  (dashed and dot-dashed curves). The symbol "$\pm$" indicates
  the sign of $\alpha_{\Lambda^*}$.}}
 \end{figure}

 In Fig.~\ref{fig:6} we show the differential
 cross sections of the $\Lambda^*$ photoproduction at $E_\gamma=2$ GeV.
 The differential cross sections of the
 $\gamma p\to \Lambda^*K^+$ reaction for positive
 $\alpha_{\Lambda^*}$ together with the separate contributions of
 the Born and $K^*$ exchange channels are shown in Fig.~\ref{fig:6}a.
 In case of the $\gamma n\to \Lambda^*K^0$ reaction, shown
 in Fig.~\ref{fig:6}b  by the solid curve, the Born term
 ($s$-channel exchange) is negligible.
 In the $\gamma p$ reaction, the interplay of the Born terms and the
 $K^*$ exchange amplitude is important at forward angles that
 leads to a dependence of the total cross section on the sign of
 $\alpha_\Lambda^*$ (see Fig.~\ref{fig:6}b).
 However, as we will see later, in the coherent
 $\gamma D\to\Lambda^*\Theta^+$ reaction the region of
 backward angles of the $K^+$ photoproduction gives the main
 contribution and, therefore, the final result is not sensitive
 to the choice of the solution. Nevertheless,
 for further consideration we chose the solution with
 positive $\alpha_{\Lambda^*}$ because it describes
 better the total $K^+K^-$ production in $\gamma p$ interaction
 at low energies.

 Finally we note that a similar approach for the $\Lambda^*$
 photoproduction based on the effective Lagrangian formalism
 was developed in the recent paper~\cite{Hosaka0503}. Differences consist in a
 different choice of the form factors and parameters, which results in
 slightly different predictions for the differential and total
 cross sections. This difference may be resolved experimentally.

\section{Reaction \boldmath{$\gamma D\to \Lambda^*\Theta^+$}}

  The tree level diagrams for the coherent
  $\gamma D\to \Lambda^*\Theta^+$ photoproduction are shown
  in Fig.~\ref{fig:7}.
\begin{figure}[t!]
{\centering
  \includegraphics[width=.42\textwidth]{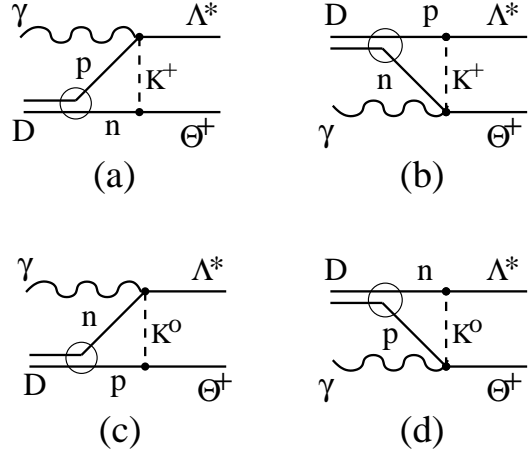}
 \caption{\label{fig:7}{\small
 Tree level diagrams for the reaction
 $\gamma D\to \Lambda^*\Theta^+$. The exchange of charged and neutral
 mesons are shown in (a,b) and (c,d), respectively.}}}
\end{figure}
First of all note that the amplitudes from the charge and neutral
meson exchange shown in Figs.~\ref{fig:7}a and c and/or b and d
give a constructive interference in the total cross section. That
is because in the elementary amplitudes of $\gamma N\to
\Lambda^*K$ and $\gamma N\to \Theta^+\bar K$ reactions the
dominant contribution comes from the $K^*$ exchange. The different
signs in $\gamma K^{0*}\bar K^0$ and $\gamma K^{+*} K^-$ vertices
are compensated by the different signs in $n\Theta^+K^-$ and
$p\Theta^+\bar K^0$ interactions. The latter is a consequence of
the assumed isospin $I=0$ of the pentaquark.

The amplitudes of the coherent $\Lambda^*\Theta^+$ photoproduction
are expressed through the transition operators of the "elementary"
processes $\gamma N\to \Lambda^*K$ and $\gamma N\to \Theta^+\bar
K$ shown in Fig.~\ref{fig:7}a,c and b,d, respectively,  as
\begin{subequations}
\label{gamma-D1}
 \begin{eqnarray}
 A_{(a,c)} &=& g_{\Theta NK}\int\frac{d^4p}{(2\pi)^4}\,
 \bar u_{\Theta}\gamma_5  \frac{1}{q^2-M_K^2}
 \bar u_{\Lambda^*}^\sigma{\cal M}^{\Lambda^*}_{\sigma\mu}
 \frac{\fs p + M}{p^2-M^2}\Gamma_D\frac{\fs p' + M}{{p'}^2-M^2}
  U_D\epsilon^\mu~,\label{gamma-D1E1}\\
 A_{(b,d)} &=&-\frac{g_{\Lambda^* NK}}{M_\Lambda^*}
 \int\frac{d^4p}{(2\pi)^4}\,
 \bar u_{\Theta}{\cal M}^{\Theta}_\mu
 \frac{1}{q^2-M_K^2}
 \bar u_{\Lambda^*}^\sigma{q}_{\sigma}\gamma_5
 \frac{\fs p + M}{p^2-M^2}\Gamma_D\frac{\fs p' + M}{{p'}^2-M^2}
  U_D\epsilon^\mu~,\label{gamma-D1E2}
\end{eqnarray}
\end{subequations}
where the transition operators ${\cal M}$ are described in the
previous section, $\Gamma_D$ and $U_D$ stand for the deuteron $np$
coupling vertex and the deuteron spinor, respectively, $p'=p_D-p$
and $q$ is the momentum of the exchanged kaon.

 Following Ref.~\cite{gammaD} we assume that the dominant
contribution to the loop integrals comes from their imaginary
parts which may be evaluated by summing  all possible cuttings of
the loops, as shown in Fig.~\ref{fig:8}.
\begin{figure}[h!]
{
  \includegraphics[width=.45\columnwidth]{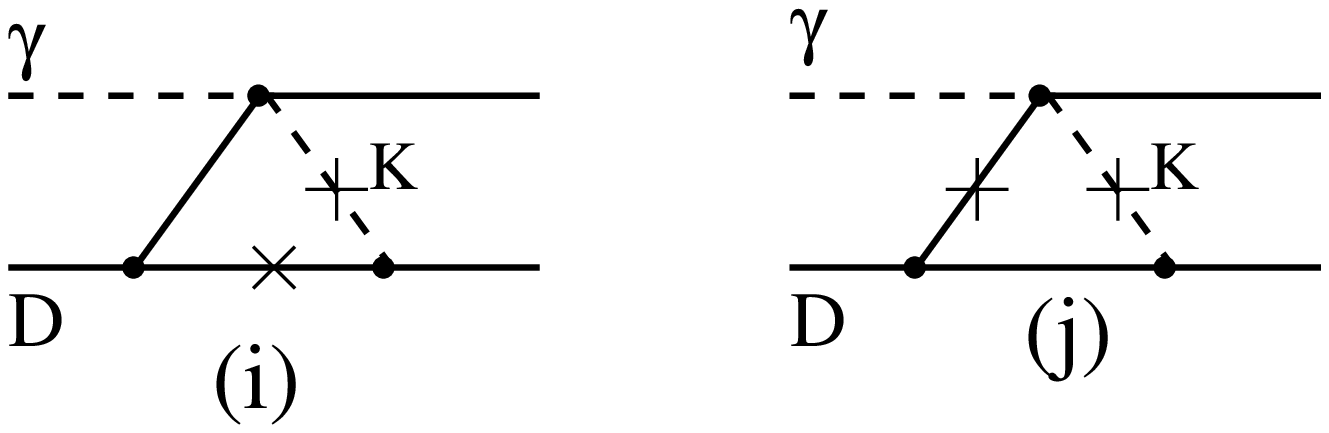}
 \caption{\label{fig:8}{\small
  Diagrammatic representation of cutting (indicated in  by crosses) the loop
  diagrams.}}}
\end{figure}
Calculating the imaginary parts we use the following substitutions
for the propagators of the on-shell particles (shown by crosses)
\begin{eqnarray}
&&\frac{1}{q^2-M_K^2}\to 2\pi\delta(q^2-M_K^2)~,\nonumber\\
&&\frac{\fs p + M}{p^2-M^2}\to 2\pi\,(\fs p + M)\,\delta(p^2-M^2)
\label{cut1}
\end{eqnarray}
and the identity
\begin{eqnarray}
\int{d^4p}\delta(p^2-M^2)=\int\frac{d^3{\bf p}}{2E} \label{cut2}
\end{eqnarray}
 with $E^2={\bf p}^2+M^2$. We also use the standard representation of the
 product of the deuteron vertex function and the attached nucleon
 propagator through the non-relativistic deuteron function
 \begin{eqnarray}
 \Gamma_D\frac{\bar u_1(p)\bar
 u_2(p_D-p)}{U_D}=\sqrt{2M_D}\,\psi_{m_D,m_1m_2}~,
 \label{cut3}
 \end{eqnarray}
 where $\psi_{m_D,m_1m_2}$ is the deuteron wave
 function with the spin projection $m_D$ and
 the nucleons spin projections $m_1$ and $m_2$.
 By using Eqs.~(\ref{cut1}) - (\ref{cut3}),
 one can express the principal parts of the invariant
 amplitudes in Eq.~(\ref{gamma-D1}) as
\begin{subequations}
\label{gamma-D2}
 \begin{eqnarray}
 A^P_{(a,c)} &=& g_{\Theta NK}
  \sum\limits_{m_1m_2}[\bar u_{\Theta}(p_\Theta)\gamma_5\, u_{m_1}(r)]
 \cdot
 [\bar u_{\Lambda^*}^\sigma(p_\Lambda^*){\cal
 M}^{\Lambda^*}_{\sigma\mu}\,\epsilon^\mu\, u_{m_2}(r)\,]\,S^{\Lambda^*}_{m_1m_2}~,
 \label{gamma-D2E1}\\
 A^P_{(b,d)} &=&-\frac{g_{\Lambda^* NK}}{M_\Lambda^*}
 \sum\limits_{m_1m_2} [\bar u_{\Theta}{\cal
 M}^{\Theta}_\mu\,u_{m_1}(r)\epsilon^\mu]
 \cdot
 [\bar
 u_{\Lambda^*}^\sigma{q}_{\sigma}\gamma_5\,u_{m_2}(r)]\,
 S^{\Theta^+}_{m_1m_2}~,\label{gamma-D2E2}
\end{eqnarray}
\end{subequations}
where $r=p_D/2$,  and
\begin{eqnarray}
 S^{\Lambda^*}_{m_1m_2}&=&I^{i}_{m_1m_2}(p_\Theta) +
 I^{j}_{m_1m_2}(k-p_\Lambda^*),\,\,\,
 S^{\Theta^+}_{m_1m_2}= I^{i}_{m_1m_2}(p_\Lambda^*) +
 I^{j}_{m_1m_2}(k-p_\Theta)~,\nonumber\\
 I^{i,j}_{m_1m_2}(p_X)&=i&\frac{\sqrt{2M_D}}{16\pi}\int\frac{p d
 p}{Ep_X}\,\theta(1-|a_{i,j}(p,p_X)|)\,\phi_{m_D,m_1m_2}(p,a(p,p_X))~,\nonumber\\
 a_{i}(p,p_X)&=&\frac{2EE_X + M_K^2 - M_X^2 - M^2}{2pp_X}~,\nonumber\\
 a_{j}(p,p_X) &=&\frac{2EE_X - M_K^2 + M_X^2 +
 M^2}{2pp_X}~,\nonumber\\
 \phi_{m_D,m_1m_2}(p,a)&=&\sqrt{4\pi}\,
 \langle\frac12 m_1\frac12 m_2|1m_D\rangle
 \left(u_0(p) +\frac{1}{\sqrt{8}}(3a^2-1)(1-3\,\delta_{m_D0})\,u_2(p)\right)~,
 \label{cut4}
 \end{eqnarray}
 where $M_X^2=E_X^2-p_X^2$ and $u_l$ with $l=0,2$ is the radial
 deuteron wave function in the momentum space, normalized as
 \begin{eqnarray}
 \int \frac{d{\bf p}}{(2\pi)^3}\,\Phi({p})=1, \nonumber
\end{eqnarray}
 where
 \begin{eqnarray}
 \Phi({p})=4\pi\left(u^2_0(p) +u^2_2(p)\right).
 \label{cut5}
 \end{eqnarray}

 \begin{figure}[h!]
 {\centering
  \includegraphics[width=.42\textwidth]{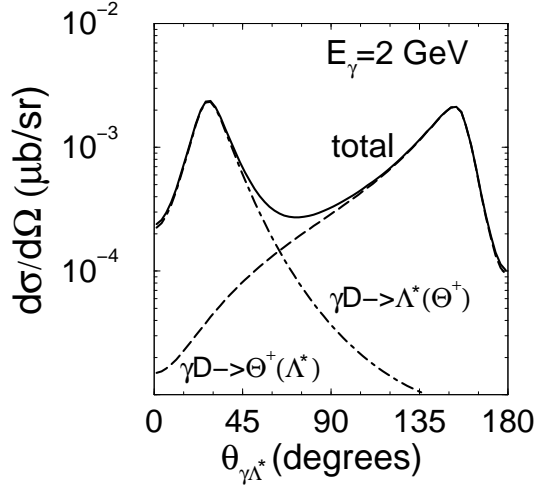}
 \caption{\label{fig:9}{\small
 The differential cross section of
 the $\gamma D\to \Lambda^*\Theta^+$ reaction.
 The notations $\gamma D\to \Lambda^*(\Theta^+)$
 and  $\gamma D\to\Theta^+(\Lambda^*)$ correspond to the
 diagrams in Fig.~\protect\ref{fig:7}a, c and b, d, respectively.}}}
\end{figure}

 In  deriving Eqs.~(\ref{gamma-D2}) we
 neglect the weak dependence of the "elementary" amplitudes of
 $\gamma N\to \Lambda^*K$ and $\gamma N\to \Theta^+\bar K$ on $p$
 (see Figs.~\ref{fig:2} and \ref{fig:4}),
 compared to the sharp $p$ dependence of $\Phi(p)$.
 In our calculation we use
 the deuteron wave function for the "realistic" Paris
 potential~\cite{Paris}. We checked that the final result does not
 depend on the fine structure of the deuteron wave
 function and practically does not depend on the choice
 of the potential.

 The differential cross section of the coherent $\Lambda^*\Theta^+$
 photoproduction reads
 \begin{eqnarray}
 \frac{d\sigma^{\gamma D\to \Lambda^*\Theta^+}}{d\Omega}
 =\frac{1}{64 \pi^2}\frac{1}{S}\frac{P_{\rm out}}{P_{\rm in}}\, |A_{a,c}+A_{b,d}|^2~,
 \label{cs-ch}
 \end{eqnarray}
 where $S, P_{\rm in}$ and $P_{\rm out}$ are the square of the total energy,
 momenta in initial and the final states in $\gamma D$ c.m.s.,
 respectively; averaging and summing over the spin projections in the initial
 and the final states are assumed.  Note that the interference between
 amplitudes $A_{a,c}$ and $A_{b,d}$ is negligible and they can be
summed  incoherently.

 In Fig.~\ref{fig:9} we show the differential cross section
 of the reaction $\gamma D\to \Lambda^*\Theta^+$ at $E_\gamma=2$~GeV
 as a function of the angle between the beam direction and direction of
 flight of $\Lambda^*$  in the $\gamma D$ c.m.s.
 The non-monotonous behaviour of the cross section is completely
 defined by the
 spectral functions $S^{\Lambda^*}$ and $S^{\Theta^+}$ in
 Eqs.~(\ref{gamma-D2E1}) and (\ref{gamma-D2E2}), respectively.
 The spectral functions $S^{\Lambda^*}$ and $S^{\Theta^+}$ have
 sharp peaks in forward
 ($\theta_{\gamma \Lambda^*}\simeq 27.5^o$) and backward
 ($\theta_{\gamma \Lambda^*}\simeq 152.5^o$ ) hemispheres,
 respectively.

\section{Background contribution}

 Since the $\Lambda^*$ and $\Theta^+$ are unstable baryons,
 the typical experiment for studying the coherent
 $\gamma D\to\Lambda^*\Theta^+$ process
 must include a simultaneous measurement of the $pK^-$ and $nK^+$
 invariant masses. Therefore, the question is whether the
 predicted cross section of the coherent $\Lambda^*\Theta^+$ photoproduction
 is large enough to be seen above the background
 of competing resonance and non-resonance processes
 in the $\gamma D\to np K^+K^-$ reaction.

 We consider three types of background processes.
 One is the photoproduction
 of a $K^+K^-$ pair in a $\gamma p$ interaction when the neutron is
 a spectator. This process includes the resonant
 $\gamma p\to \Lambda^*K^+\to pK^+K^-$ photoproduction
 and the non-resonant $\gamma p\to pK^+K^-$
 reaction shown in Fig.~\ref{fig:10}a and b, respectively.
 \begin{figure}[h!]
 {\centering
  \includegraphics[width=.42\textwidth]{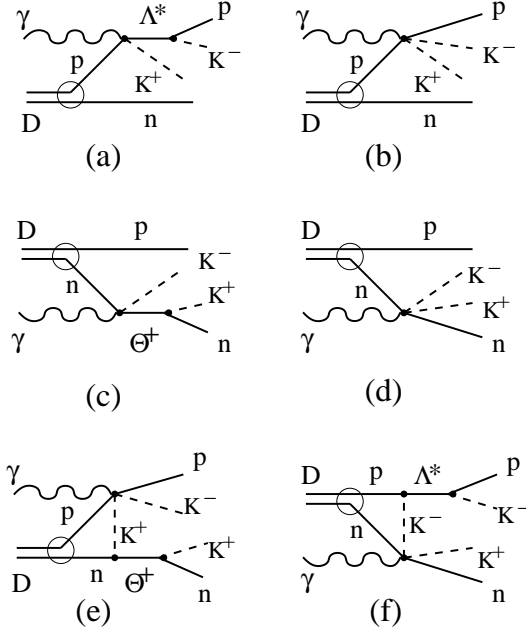}
 \caption{\label{fig:10}{\small
 Tree level diagrams for background processes.
 a - d: non-coherent spectator channels, e and f: coherent
  semi-resonant background processes.}}}
 \end{figure}

 Similarly, a $K^+K^-$ pair can be produced in a $\gamma n$
 interaction, when the proton is a spectator.
 The corresponding processes are depicted in  Fig.~\ref{fig:10}c and d.

 The third process is the coherent "background" when the $K^+K^-$ pair
 is produced in a $\gamma N$ interaction and one of the kaons
 together with the second nucleon forms the outgoing
 $\Theta^+$ or $\Lambda^*$, as shown in
 Fig.~\ref{fig:10}~e and f, respectively. We denote it as a
 coherent semi-resonant background.

 \subsection{Spectator channels}

 First, let us consider the $K^+K^-$ photoproduction in a $\gamma D$
 interaction where the neutron or proton are merely spectators.
 As an input, we have to describe the elementary processes
 $\gamma p\to p K^+K^-$ and $\gamma n\to n K^+K^-$ which
 consist of the resonant and non-resonant parts.

 \subsubsection{\boldmath$\gamma p\to pK^+K^-$}

 The dominant contribution to the
 non-resonant part in $\gamma p$ reactions comes from the virtual
 vector meson decay and $\Lambda(1405)$
 excitation~\cite{TEHN04,Oh:2004wp} as
 depicted in Fig.~\ref{fig:11}a and b.
 The contribution from excitations of other hyperons is strongly
 suppressed since they are far off-shell.
\begin{figure}[h!]
 {\centering
  \includegraphics[width=.42\textwidth]{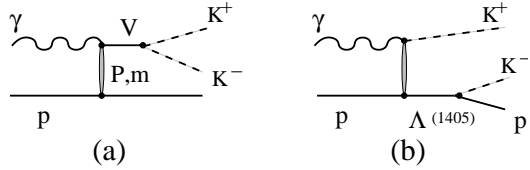}
 \caption{\label{fig:11}{\small
  Background processes for the $\gamma p\to p K^+K^-$ reaction.
  (a): vector meson contribution. (b): virtual $\Lambda(1405)$
  excitation.}}}
 \end{figure}
The vector meson channel $\gamma p\to V p\to pK^+K^-$, where
$V=\phi,\rho,\omega$ has been analyzed in detail in
 Ref.~\cite{TEHN04}. In the present study we use this model where
 the vector mesons are produced through the Pomeron and meson
 ($\pi,\eta,\sigma$) exchanges with the same parameters.
The only difference with Ref.~\cite{TEHN04} is that now we do not
use a cut on the invariant mass of the $K^+K^-$ pair around the
$\phi$ meson mass.

We parameterize the amplitude of the virtual $\Lambda(1405)$
excitation through the $K^*$ exchange process. This assumption is
supported by the $K^*$ exchange dominance in $\Lambda^*$ and
$\Theta^+$ photoproduction and allows to reduce the number of
unknown parameters. The amplitude of this channel reads
\begin{eqnarray}
 A^{\Lambda'}_{fi}&=&\bar u(p')\,
 {\cal M}_{\mu}^{\Lambda'}\,u(p)\, \varepsilon^{\mu}~,\nonumber\\
 {\cal M}_{\mu}^{\Lambda'}&=&-i\frac{eg_{\gamma KK^*}g'}
 {M_{K^*}(t-M_{K^*}^2)}
 \varepsilon^{\mu\nu\alpha\beta}k_\nu q_\alpha\,
 \frac{(\fs p_{\Lambda' + M_{\Lambda'}})\gamma_5\gamma_\beta}
 {p^2_{\Lambda'} - M_{\Lambda'}^2 +i
 \Gamma_{\Lambda'}M_{\Lambda'}}\,F_{K^*}(t)~,
\end{eqnarray}
 where $\Lambda'\equiv \Lambda(1405)$, $\Gamma_{\Lambda'}=50$ MeV
 is the total decay width of $\Lambda'$~\cite{PDG},
 $F_{K^*}(t)$ is the $K^*$ exchange form factor, the constant $g'$ is a
 product of two coupling constants $g_{\Lambda' NK}$ and $g_{\Lambda' NK^*}$.
 The choice $g'\simeq7.8$ gives the correct value of the total yield of
 $K^+K^-$ mesons at $E_\gamma\sim 2$ GeV. Note that the interference
 between the resonance and non-resonance channels
 in the total cross section
 is rather weak and, therefore, they can be added incoherently.
 Thus, the total cross section of the $\gamma p\to p K^+K^-$
 reaction reads
\begin{eqnarray}
\frac{d\sigma}{d\Omega dM_{pK^-}}&=&
\left(\frac{d\sigma}{d\Omega}\right)^{\gamma
 p\to\Lambda^*K^+}\,F^{\Lambda^*}(M_{pK^-})\nonumber\\
 &+&\frac{1}{64\pi^2}\frac{1}{s}\frac{p_{\rm out}}{p_{\rm in}}\frac{\bar q_F}{16\pi^3}
 \int\left( |A^{V}_{fi}(\gamma p)|^2 +
 |A^{\Lambda'}_{fi}|^2\right)\,d\Omega_F~,
\label{gp-bg}
\end{eqnarray}
 where $\Omega$ is the solid angle of the $K^-$ meson photoproduction
 in the $\gamma p$ c.m.s., $\bar q_F$ is
 the momentum of the $K^-$ meson in the c.m.s. of the
 $pK^-$-pair, $\Omega_F$ is the  $K^-$ meson solid angle in this
 system. Summing and averaging over the spin projection in
 the initial and the final states is to be included.
  $F^{\Lambda^*}(M_{pK^-})$ stands
 for the $\Lambda^*$ decay distribution which is obtained
 straightforwardly  from the general expression of the
 $\gamma p\to pK^+K^-$ amplitude with the virtual excitation of
 a $\Lambda^*$~hyperon,
\begin{eqnarray}
 F^{\Lambda^*}(M_x)=\frac{\Gamma_{\Lambda^*\to pK^-}}{\pi\Gamma_{\rm tot}}
 \frac{2M_xM_{\Lambda^*}\Gamma_{\rm tot}}
 {(M_x^2-M_{\Lambda^*}^2)^2 + (\Gamma_{\rm tot}M_{\Lambda^*})^2},
 \label{f-lambda}
\end{eqnarray}
 where $\Gamma_{\rm tot}=15.6$ MeV and
 $\Gamma_{\Lambda^*\to p K^-}=(0.45/2)
 \times \Gamma_{\rm tot}$~\cite{PDG}.

\begin{figure}[h!]
 {\centering
 \includegraphics[width=.42\textwidth]{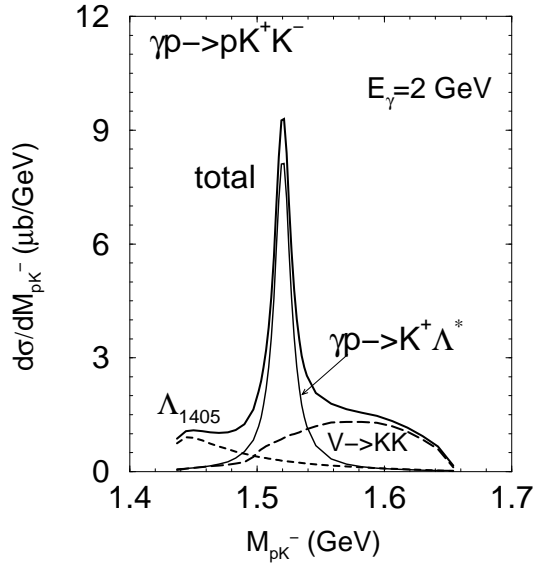}
 \caption{\label{fig:12}{\small
 The $pK^-$ invariant mass distribution in the $\gamma p\to pK^+K^-$
 reaction at $E_\gamma=2$ GeV.
 The resonant channel, vector meson and $\Lambda(1405)$ contributions are
 shown by thin solid, long dashed and dashed curves, respectively.}}}
 \end{figure}

 The $pK^-$ invariant mass distribution at
 $E_\gamma=2$ GeV integrated over $\Omega$ is shown in
 Fig.~\ref{fig:12}. One can see that the $\Lambda(1405)$ excitation
 contributes at $M_{pK^-}$ below the $\Lambda^*$ resonance
 position, and the vector meson channels contribute mainly at
 large $M_{pK^-}$, above $M_{\Lambda^*}$. The partial
 contributions
 to the total $\gamma p\to pK^+K^-$ cross section are the
 following: $\sigma(\Lambda^*)\simeq0.19\,\mu$b,
 $\sigma(V)\simeq0.17\,\mu$b and
 $\sigma(\Lambda(1405))\simeq0.07\,\mu$b. The total cross section
 $\sigma_{\rm tot}\simeq0.43\mu$b is in agreement with the
 experimental data of Ref.~\cite{exp-gp-KK}:
 $\sigma^{\rm exp}_{\rm tot}=(0.47\pm0.12)\,\mu$b at
 $E_{\gamma}=2-2.5$~(GeV).

\subsubsection{\boldmath $\gamma n\to n K^+K^-$}

In this case the non-resonance part is dominated by the vector
meson excitation and, therefore, the $nK^+$ invariant mass
distribution may be written in obvious notation as
\begin{eqnarray}
\frac{d\sigma}{d\Omega dM_{nK^+}}&=&
\left(\frac{d\sigma}{d\Omega}\right)^{\gamma
 n\to\Theta^+K^-}\,F^{\Theta^+}(M_{nK^+})
 + \frac{1}{64\pi^2}\frac{1}{s}\frac{p_{\rm out}}{p_{\rm in}}\frac{q_F}{16\pi^3}
 \int |A^{V}_{fi}(\gamma n)|^2\,d\Omega_F
 \label{gn-bg}
\end{eqnarray}
with
\begin{eqnarray}
 F^{\Theta}(M_x)=\frac{1}{2\pi}
 \frac{2M_xM_{\Theta}\Gamma_{\Theta}}
 {(M_x^2-M_{\Theta}^2)^2 + (\Gamma_{\rm tot}M_{\Theta})^2}.
\label{theta-bw}
\end{eqnarray}
We will also use the Gaussian distribution taking into account the
small $\Theta^+$ decay width and the finite experimental
resolution
\begin{eqnarray}
 F^{\Theta}_G(M_x)=\frac{1}{2}\frac{1}{\sigma\sqrt{2\pi}}
 e^{-\frac{(M_x-M_\Theta)^2}{2\sigma^2}}.
 \label{gauss}
\end{eqnarray}

\begin{figure}[h!]
 {\centering
 \includegraphics[width=.42\textwidth]{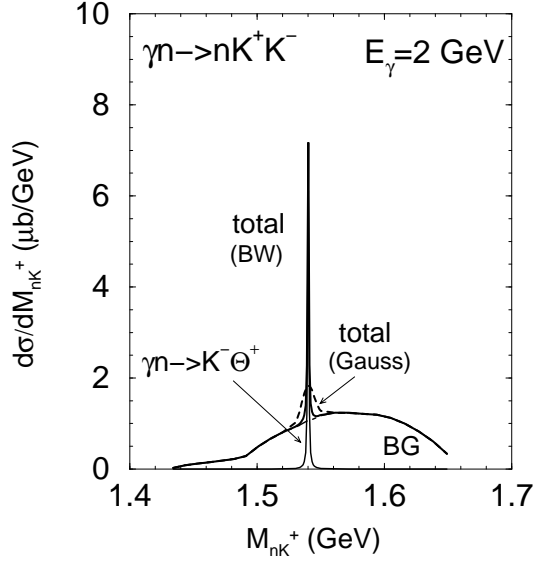}
 \caption{\label{fig:13}{\small
 The $nK^+$ invariant mass distribution in the $\gamma n\to nK^+K^-$
 reaction at $E_\gamma=2$ GeV. The symbol "BW" means the Breit-Wigner
 $\Theta^+$ decay distribution of Eq.~(\protect\ref{theta-bw}).
 The dashed curve corresponds to a Gaussian
 distribution of the $\Theta^+$ decay width $\sigma=5$ MeV.}}}
 \end{figure}

 The $nK^+$ invariant mass distribution at
 $E_\gamma=2$ GeV integrated over $\Omega$ is shown in
 Fig.~\ref{fig:13}. One can see the sharp peak
 of $\Theta^+$ excitation. In  case of a Gaussian
 $\Theta^+$ decay distribution the peak is modified.
 The height of the peak is reduced by the factor
 $\sigma/\Gamma_\Theta$ and the width becomes proportional to $\sigma$.

\subsubsection{\boldmath Spectator reactions $\boldmath{\gamma D\to pK^+K^-(n)}$
and $\boldmath{\gamma D\to nK^+K^-(p)}$}

The differential cross section of the $\gamma D\to pK^+K^-(n)$
reaction, where the neutron is a spectator, reads
\begin{eqnarray}
 &&\frac{d\sigma^{\rm sp.(n)}}{d\Omega dM_{pK^-}dM_{nK+}}=
 \left(\frac{d\sigma}{d\Omega dM_{pK^-}}\right)^{\gamma p\to
 pK^+K^-}\,W_{nK}(M_{nK^+})~,\nonumber\\
 && W_{nK}(M_{nK^+})= 2M_{nK^+}\,
 \int \frac{d{\bf p}_n}{(2\pi)^3\sqrt{1 +{\bf p}_n^2/M_N^2}}
 \,\delta(M_{nK^+}^2 -(p_n + q)^2)\,\,\Phi({\bf p}_n)~,
\label{sp-n1}
\end{eqnarray}
 where we neglect the smooth dependence of ${d\sigma}^{\gamma p\to pK^+K^-}$
 on ${\bf p}_n$ in comparison to the sharp ${\bf p}_n$ dependence of the
 momentum distribution in the deuteron, $\Phi({\bf p}_n)$,
 defined in Eq.~(\ref{cut5}).

 If the invariant mass of the $nK^+$ pair is not fixed then the
 integration over $M_{nK+}$ leads to the obvious result
\begin{eqnarray}
 \int dM_{nK^+}\,\frac{d\sigma^{\rm sp.(n)}}{d\Omega dM_{pK^-}dM_{nK+}}\simeq
 \left(\frac{d\sigma}{d\Omega dM_{pK^-}}\right)^{\gamma p\to
  pK^+K^-}~.
\label{sp-n2}
\end{eqnarray}
When the invariant mass is fixed then the function
$W_{nK}(M_{nK^+})$ becomes important and, moreover, it mainly
defines the dependence of the cross section on $M_{nK^+}$. Indeed,
let us assume that the momentum distribution in a deuteron behaves
like a delta function, i.e.  $\Phi({\bf p})\simeq
(2\pi)^3\,\delta({\bf p})$. Then one gets
\begin{eqnarray}
 W_{nK}(M_{nK^+})\simeq
 \,2M_{nK^+}\,\delta(M_{nK^+}^2 -(M_N^2+M_K^2 + 2E_{K^+}M_N)).
\label{sp-n3}
\end{eqnarray}
 That is, the distribution  $W_{nK}(M_{nK^+})$ has a peak around
 the point ${M_{nK^+}}_0\simeq \sqrt{M_N^2+M_K^2 + 2E_{K^+}M_N}$ which
 is determined by the energy of the $K^+$ meson in the laboratory system.
 On the other hand, this energy depends on the invariant mass
 of the $pK^-$ pair and the angle of the $K^+$ production in the
 $\gamma p$ c.m.s. In reality, the distribution function reads
\begin{eqnarray}
 W_{nK}(M_{nK^+})&=& 2M_{nK^+}\int \frac{p d{p}}{8\pi^2q_L\sqrt{1 +{p}^2/M_N^2}}
 \,\Phi(p)\,\theta(1-|a|)~,\nonumber\\
 a&=&\frac{2\sqrt{(q_L^2+M_K^2)(p^2+M_N^2)} + M_N^2  +  M_K^2 -
 M_{nK^+}^2}{2pq_L}~,
 \label{sp-n4}
\end{eqnarray}
 where $q_L$ is the momentum of $K^+$ meson in laboratory system.
 The distribution function $W_{nK}$ is shown
 in Fig.~\ref{fig:14}a as a function of $M_{nK^+}$
 at fixed angle of $pK^-$ pair photoproduction, $\theta_{\gamma (pK^-)}$
 (in $\gamma D$ c.m.s.) for  three different
 invariant masses of the $pK^-$ pair: $M_{pK^-}=$1.52, 1.57 and
 1.47 GeV. The choice of $\theta_{\gamma (pK^-)}=27.5^o$
 corresponds to the position of the maximum of the coherent
 $\gamma D\to \Lambda^*\Theta^+$ photoproduction cross section
 at forward angles (see Fig.~\ref{fig:9}). This angle corresponds to
 the backward $K^+$ photoproduction in $\gamma p\to\Lambda^*K^+$:
 $\theta_{\gamma K^+}\simeq119^o$ in the $\gamma p$ c.m.s.

 \begin{figure}[h!]
 {\centering
 \includegraphics[width=.4\textwidth]{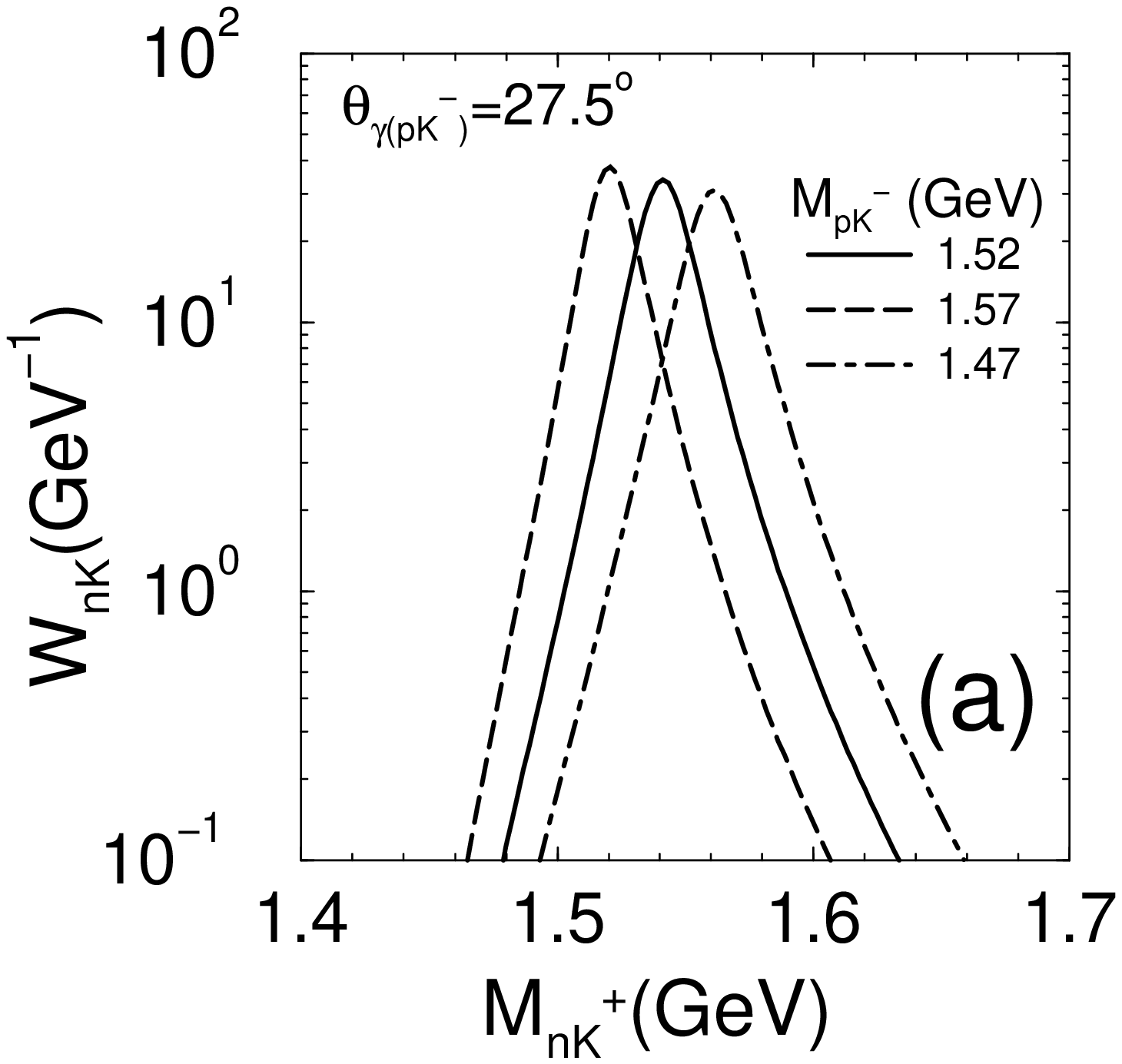}\qquad
\includegraphics[width=.4\textwidth]{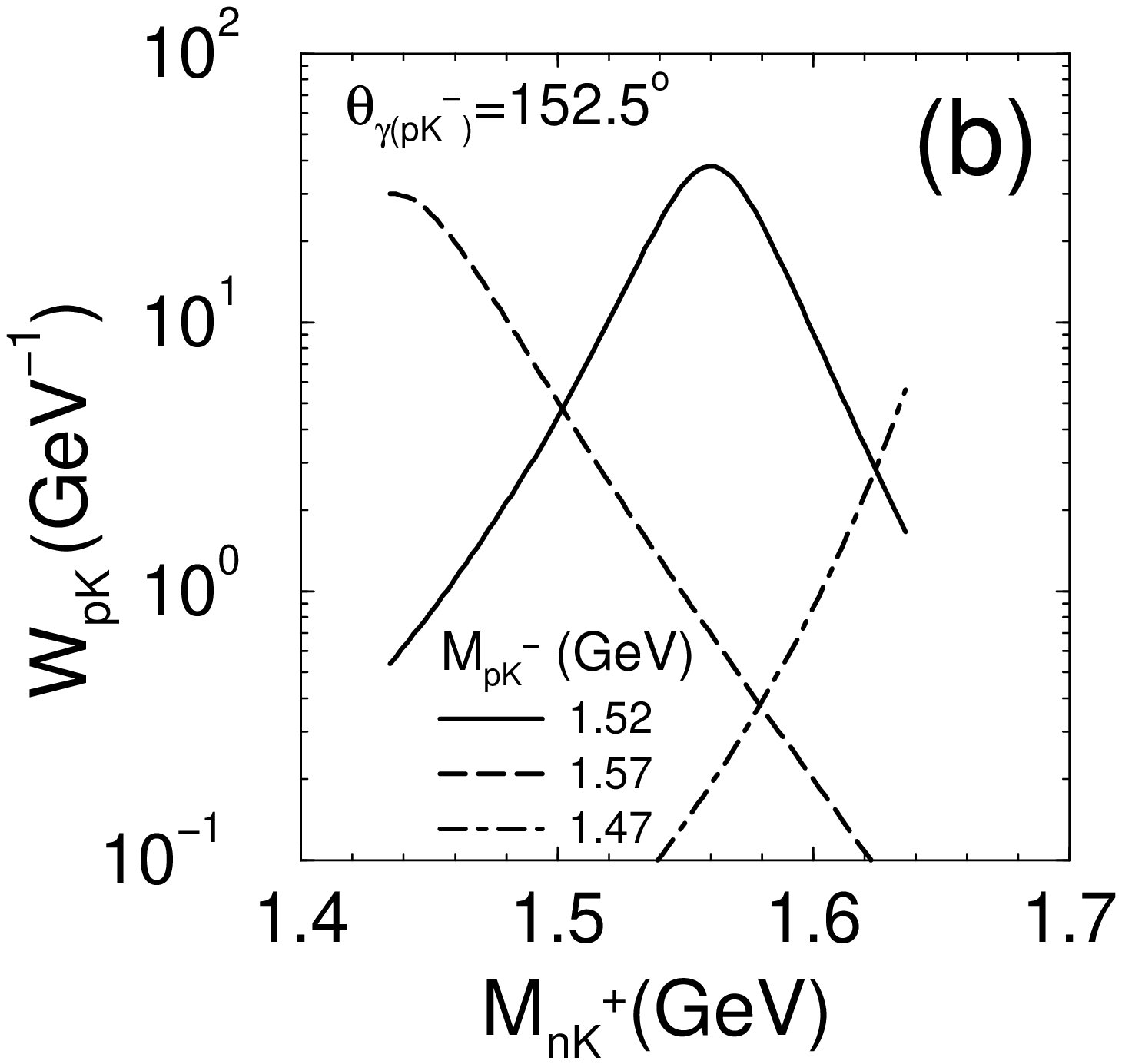}
 \caption{\label{fig:14}{\small
 (a) The invariant mass distribution function $W_{nK}$
 as a function of $M_{nK^+}$ at $\theta_{\gamma (pK^-)}=27.5^0$ and
 fixed values of $M_{pK^-}$.
 (b) The invariant mass distribution function $W_{pK}$
 as a function of $M_{nK^+}$ at $\theta_{\gamma (pK^-)}=152.5^0$ and
 fixed values of $M_{pK^-}$. }}}
 \end{figure}

 The differential cross section of the $\gamma D\to nK^+K^-(p)$
 reaction, where the proton is spectator, may be obtained from
 Eq.~(\ref{sp-n1}), using the substitution $n\to p$, $K^+\to K^-$
 and $M_{nK^+}\to M_{pK^-}$,
\begin{eqnarray}
 &&\frac{d\sigma^{\rm sp.(p)}}{d\Omega dM_{pK^-}dM_{nK+}}=
 \left(\frac{d\sigma}{d\Omega dM_{nK^+}}\right)^{\gamma n\to
 nK^+K^-}\,W_{pK}(M_{pK^-})~.
\label{sp-p1}
\end{eqnarray}
 The essential difference is  that now we analyze the dependence of
 the distribution function $W_{pK}$ not on $M_{pK^-}$ but
 on the invariant mass $M_{nK^+}$. This dependence is included in
 $W_{pK}$ implicitly through the dependence of the momentum of $K^-$
 on $M_{nK^+}$ and therefore, in general, we have no narrow
 peak structure of  $W_{pK}$ as a function of  $M_{nK^+}$.
 As an example, in Fig.~\ref{fig:14}b we show the distribution
 $W_{pK}$ as a function of  $M_{nK^+}$ at fixed values of
 $M_{pK^-}=$1.52, 1.57 and 1.47 GeV and $\theta_{\gamma
 (pK^-)}=152.5^o$. One can see a broad maximum at $M_{pK^-}=$1.52 GeV
 and an almost monotonic behaviour at 1.47 and 1.57 GeV.

\subsection{Coherent semi-resonant background}

The amplitude of the process shown in Fig.~\ref{fig:10}e is
calculated similarly to the amplitude of the coherent
$\Lambda^*\Theta^+$ photoproduction described by
Eq.~(\ref{gamma-D2E1}). The corresponding cross section reads
\begin{eqnarray}
&&\frac{d\sigma^{e}}{d\Omega dM_{pK^-}dM_{nK+}}
 =\frac{1}{64\pi^2}\frac{1}{s}\frac{p_{\rm out}}{p_{\rm in}}
 \frac{\bar q_F}{16\pi^3}\
 \frac{1}{2}\Big|\frac{d\Omega'}{d\Omega}\Big|
 \int\,d\Omega_F\,|A_e|^2
 \,F^\Theta(M_{nK^+})\,
 ~,\nonumber\\
&&A_e= g_{\Theta NK}
  \sum\limits_{m_1m_2}[\bar u_{\Theta}(p_\Theta)\gamma_5\, u_{m_1}(r)]
 \cdot
 [\bar u_{\Lambda^*}^\sigma(p_\Lambda^*){\cal
 M}^{\gamma p\to pK^+K^-}_{\sigma\mu}\,
 \varepsilon^\mu\, u_{m_2}(r)\,]\,S^{\Lambda^*}_{m_1m_2}~,
 \label{ch-1}
\end{eqnarray}
where $p_{\rm in}$, $p_{\rm out}$ are the momenta of the proton
and $pK^-$ pair in $\gamma p$ c.m.s., $\Omega$ and $\Omega'$  are
the solid angles of the $pK^-$ pair in $\gamma D$ and $\gamma p$
reactions, respectively, $\bar q_F$ is the momentum of $K^-$ meson
in the rest frame of the $pK^-$ pair, $\Omega_F$ is the solid
angle of $K^-$ in this frame. The additional factor $1/2$ assumes
renormalization  of the flux in the $\gamma D$ system compared to
the $\gamma p$ interaction. The function $F^\Theta(M_{nK^+})$ is
defined in Eq.~(\ref{theta-bw}).
 Averaging and summing over the spin projections in
initial and the final states, respectively, have to be performed.
Actually, here we have a sum of two cross sections. One is the
contribution of the virtual vector meson and another one is the
contribution of the virtual $\Lambda(1405)$ excitation.

Similarly, one can write the cross section of the process shown in
Fig.~\ref{fig:10}f as
\begin{eqnarray}
&&\frac{d\sigma^{f}}{d\Omega dM_{pK^-}dM_{nK+}}
 =\frac{1}{64\pi^2}\frac{1}{s}\frac{p_{\rm out}}{p_{\rm in}}\,
 \frac{q_F}{16\pi^3}\,
 \frac{1}{2}\Big|\frac{d\Omega'}{d\Omega}\Big|
 \int\,d\Omega_F |A_f|^2\,F^{\Lambda^*}(M_{pK^-})~,\nonumber\\
&&A_f= -\frac{g_{\Lambda^* NK}}{M_\Lambda^*}
 \sum\limits_{m_1m_2} [\bar u_{\Theta}{\cal
 M}^{\gamma n\to nK^+K^-}_\mu\,u_{m_1}(r)\varepsilon^\mu]
 \cdot [\bar
 u_{\Lambda^*}^\sigma{q}_{\sigma}\gamma_5\,u_{m_2}(r)]\,
 S^{\Theta^+}_{m_1m_2}~,
 \label{ch-2}
\end{eqnarray}
 where the function $F^{\Lambda^*}(M_{pK^-})$ is defined in
 Eq.~(\ref{f-lambda}) and other notations are similar to the
 previous case.
\begin{figure}[h!]
 {\centering
 \includegraphics[width=.4\textwidth]{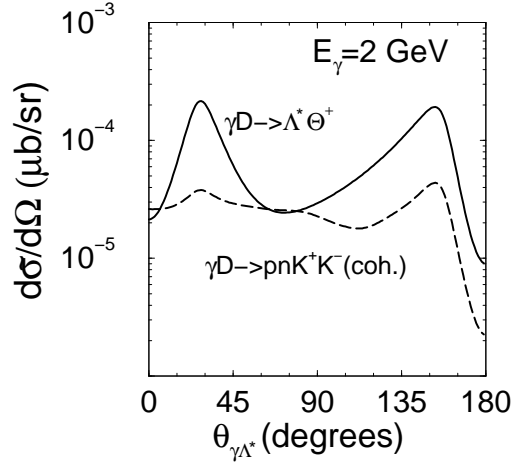}
 \caption{\label{fig:15}{\small
 Comparison of the coherent $\Lambda^*\Theta^+$ photoproduction
 (solid curve) and
 coherent semi-resonant background (dashed curve) depicted in
 Fig.~\protect\ref{fig:10}e,f.}}}
 \end{figure}

 Let us now compare the contribution of the coherent $\Lambda^*\Theta^+$
 photoproduction and the coherent semi-resonant background described by
 Eqs.~(\ref{ch-1}) and (\ref{ch-2}) in the vicinity of the $\Theta^+$
 and $\Lambda^*$ resonance position
 \begin{eqnarray}
 &&
 \frac{d\widetilde{\sigma}^{ch.}}{d\Omega}=
 \int\limits_{M_{\Lambda^*}-\Delta}^{M_{\Lambda^*}+\Delta}
\, \int\limits_{M_{\Theta}-\Delta}^{M_{\Theta}+\Delta}
 dM_{pK^-}\,dM_{nK^+}\frac{d{\sigma}^{\gamma D\to
\Lambda^*\Theta^+}}{d\Omega}
\,F_{\Lambda^*}(M_{pK^-})\,F_{\Theta^+}(M_{nK^+})~,\nonumber\\
&& \frac{d\widetilde{\sigma}^{ch.bg.}}{d\Omega}=
 \int\limits_{M_{\Lambda^*}-\Delta}^{M_{\Lambda^*}+\Delta}
 \int\limits_{M_{\Theta}-\Delta}^{M_{\Theta}+\Delta}
  dM_{pK^-}\,dM_{nK^+}
 \left(\frac{d\sigma^{e}}{d\Omega dM_{pK^-}dM_{nK+}}+
 \frac{d\sigma^{f}}{d\Omega dM_{pK^-}dM_{nK+}}\right),
\label{cs-ch-nch}
 \end{eqnarray}
 where $\Delta=20$ MeV. In Fig.~\ref{fig:15} we show result of such
 a comparison. One can see that the coherent background
 contribution has
 local maxima caused by the spectral functions $S$, but the
 values of these contributions at the peak positions are much smaller
 compared to the coherent process. Therefore, the dominant
 background contribution comes from the spectator processes.

\section{Results and discussion}

 As pointed out above, the coherent $\Lambda^*\Theta^+$
 photoproduction seems to be accessible  most effectively by a search
 for a sharp $\Theta^+$ peak in the invariant $nK^+$ mass distribution at
 fixed invariant masses of the $pK^-$ pair
 \begin{eqnarray}
 \frac{d{\sigma}^{\gamma D\to npK^+K^-}(M_0)}{d\Omega dM_{nK^+}}=
 \int\limits_{M_{0}-\Delta}^{M_{0}+\Delta}
 \, dM_{pK^-}\,
 \frac{d{\sigma}^{\gamma D\to npK^+K^-}}{d\Omega
 dM_{nK^+}dM_{pK^-}}~.
 \end{eqnarray}
 In our further analysis we choose $M_0=1.52,\,1.57$, and $1.47$ GeV
 and $\Delta=20$~MeV. One can expect that the coherent
 photoproduction appears at $M_0=M_{\Lambda^*}=1.52$ GeV and it
 is suppressed relative to the strong background  when we go above or below
 this point. Since the cross section of the coherent
 photoproduction at $E_\gamma=2$ GeV has  bumps at
 $\theta_{\gamma\Lambda^*}\simeq27.5^o$ and $152.5^o$ in $\gamma
 D$ c.m.s. (see Fig.~\ref{fig:9}), then it is natural to expect that
 the regions around these angles are more favored  for a manifestation
 of the coherence effect.

\begin{figure}[h!]
 {\centering
 \includegraphics[width=.3\textwidth]{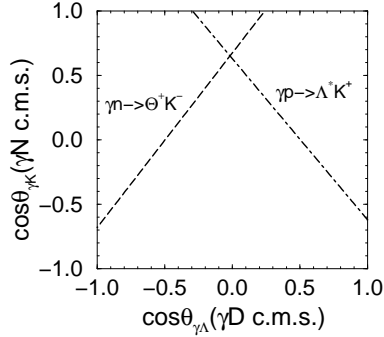}
 \caption{\label{fig:16}{\small
 The dependence of  $\cos\theta_{\gamma K^-}$ in $\gamma p\to \Lambda^*K^+$
 photoproduction and $\cos\theta_{\gamma K^+}$ in $\gamma n\to \Theta^+K^-$
 photoproduction as a function of $\cos\theta_{\gamma\Lambda^*}$ in
 the $\gamma D\to \Lambda^*\Theta^+$ reaction at $E_\gamma=2$ GeV.}}}
 \end{figure}

 Note that at forward  and backward angles
 of the $pK^-$ pair photoproduction, $\theta_{\gamma(pK^-)}$, some of
 spectator processes shown in Fig.~\ref{fig:10} are suppressed
 dynamically. To illustrate this point let us consider
 the dependence of $\cos\theta_{\gamma K^-}$ in $\gamma p\to \Lambda^*K^+$
 photoproduction and $\cos\theta_{\gamma K^+}$ in $\gamma n\to \Theta^+K^-$
 photoproduction as a function of $\cos\theta_{\gamma\Lambda^*}$. Here we assume
 that $\theta_{\gamma K}$ is the $K$ meson photoproduction angle in the $\gamma N$
 c.m.s. and $\theta_{\gamma\Lambda^*}$ is the $\Lambda^*$
 photoproduction angle in $\gamma D$ c.m.s.
 One can see that the region of $0\leq\theta_{\gamma\Lambda^*}\lesssim 76^o$
 is forbidden kinematically for  $\Theta^+K^-$ photoproduction
 from the resting neutron. Similarly, the region of $107^o\lesssim\theta_{\gamma\Lambda^*}\leq\pi$
 is forbidden for  $\Lambda^*K^+$ photoproduction from the resting proton.
 In the kinematically forbidden regions the corresponding
 processes can be proceeded only through the high-momentum component in the
 deuteron wave function and, therefore, are exponentially small.

 Consider first $\gamma D\to npK^+K^-$ photoproduction at
 a  forward angle of the $pK^-$ pair
 at $\theta_{\gamma(pK^-)}\simeq27.5^o$ and $E_\gamma=2$ GeV. The corresponding
 invariant mass distributions for $M_0=1.52$, 1.57 and 1.47 are shown
 in Fig.~\ref{fig:17}a, b and c, respectively.
\begin{figure}[h!]
 {\centering
 \includegraphics[width=.3\textwidth]{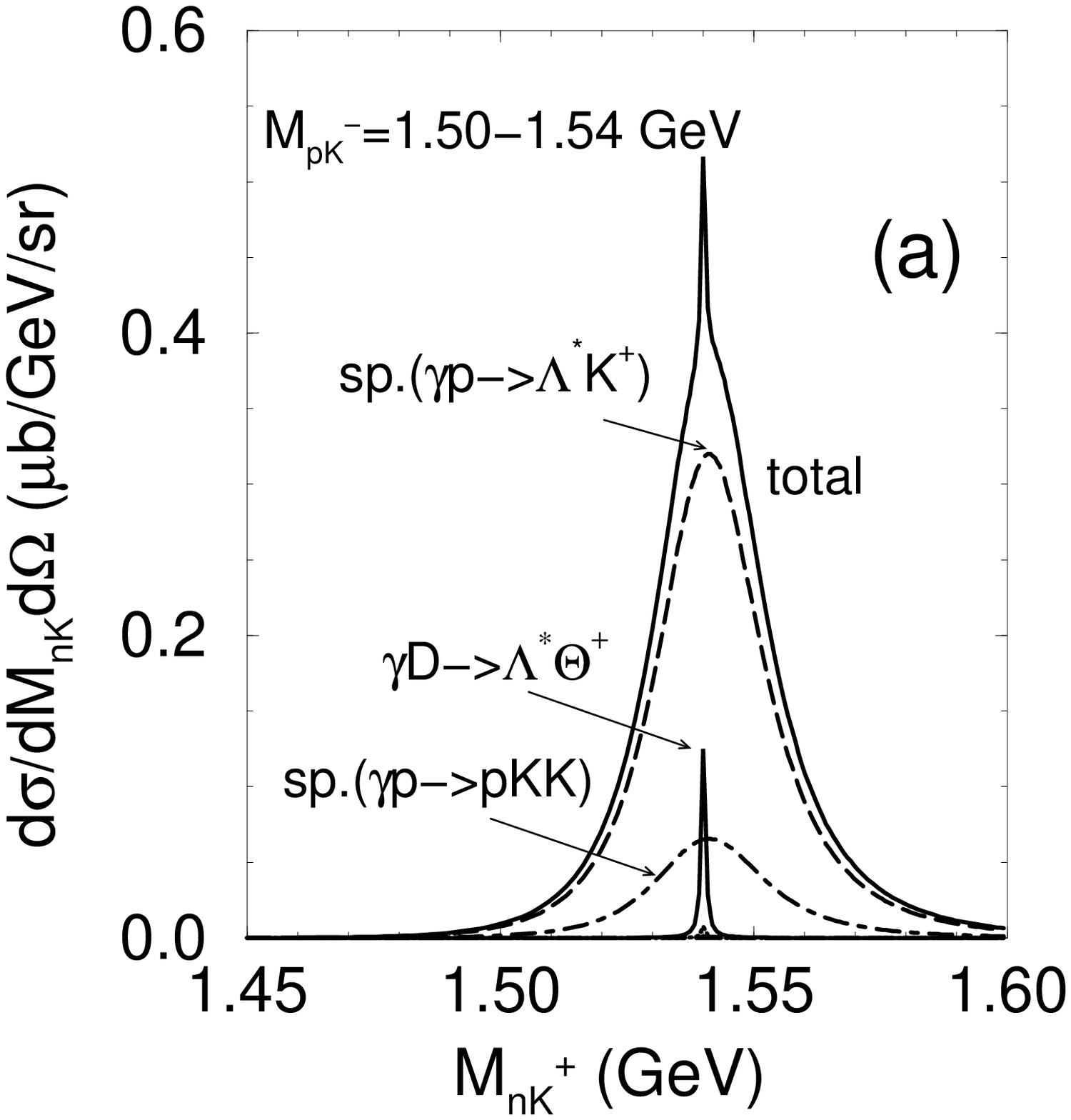}\qquad
 \includegraphics[width=.3\textwidth]{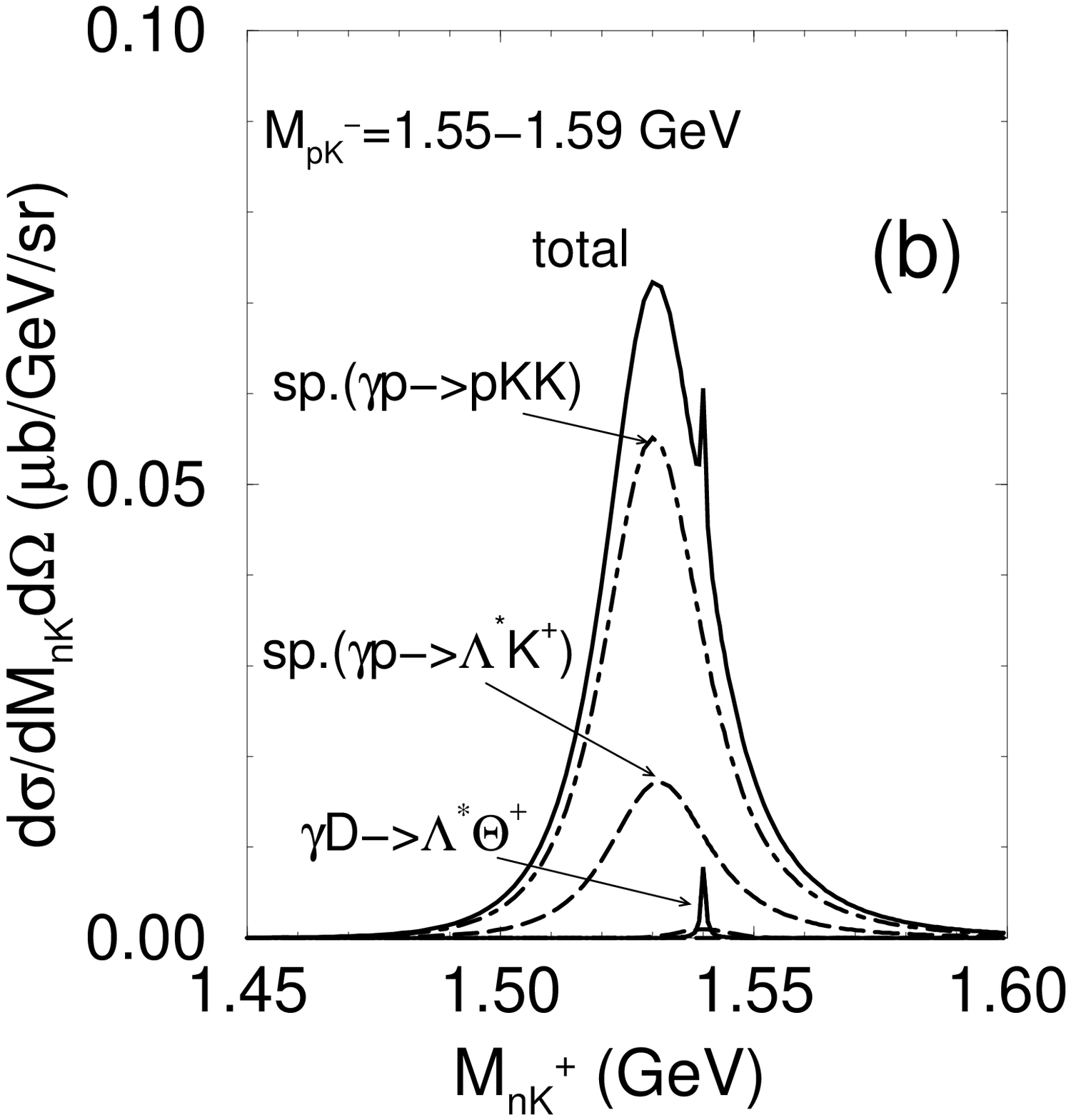}\qquad
 \includegraphics[width=.3\textwidth]{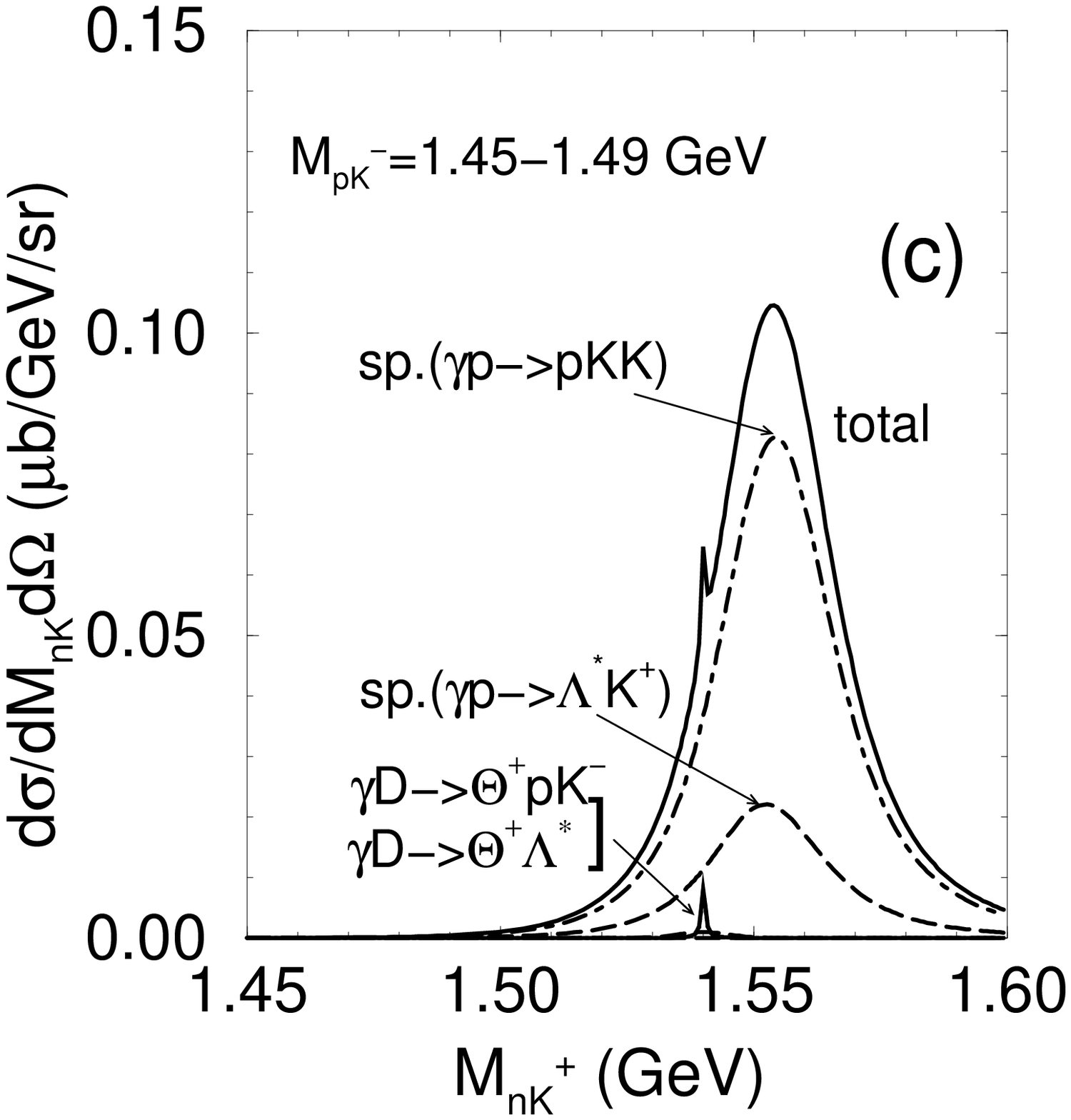}\qquad
 \caption{\label{fig:17}{\small
 The $nK^+$ invariant mass  distribution in the $\gamma D\to npK^+K^-$
 reaction at fixed values of the $pK^-$ invariant mass.
 The angle of the  $pK^-$ pair photoproduction
 in $\gamma D$ c.m.s., $\theta_{\gamma(pK^-)}=27.5^o$ and  $E_\gamma = 2$ GeV.
 (a) $M_{pK^-}=1.52\pm0.02$ GeV; (b) $M_{pK^-}=1.57\pm0.02$ GeV,
 (c) $M_{pK^-}=1.47\pm0.02$ GeV. Notations "sp.$(\gamma p\to \Lambda^*K^+)$"
 and "sp.$(\gamma p\to pKK)$" correspond to the processes
 depicted in Fig.~\protect\ref{fig:10}a and b, respectively;
 "$\gamma D\to \Theta^+pK^-$" corresponds to the coherent background
 shown in Fig.~\protect\ref{fig:10}e, "$\gamma D\to\Lambda^*\Theta^+$" corresponds
 to the coherent $\Lambda^*\Theta^+$ photoproduction
 (Fig.~\protect\ref{fig:7}).}}}
 \end{figure}

 At $M_0=M_{\Lambda^*}$, the background is dominated by the
 resonant $\Lambda^*$ photoproduction in the spectator mechanism
 shown in Fig.~\ref{fig:10}a. The next important contribution comes
 from the non-resonant spectator
 channel (Fig.~\ref{fig:10}b). The shape of the background spectrum
 has a resonance
 like behavior with the center close to the mass of
$\Theta^+$
 and a width of about 15~MeV. This behaviour is defined by the
 spectral distribution function $W_{nK}$ (or the deuteron momentum
 distribution) in Eq.~(\ref{sp-n1})
 and the kinematics (see Fig.~\ref{fig:14}a). At
 $M_{pK^-}=1.52$~GeV,
 $W_{nK}$ has a sharp peak at $M_{nK^+}\simeq1.54$ GeV.
 For $M_{pK^-}=$1.57 and 1.47 GeV  the peak position is  shifted
 to lower  or higher masses, respectively.
  Similarly, one can see the corresponding
 shift in the background contribution at $M_0=1.57$ and 1.47 GeV,
 shown in Figs.~\ref{fig:17}b and c. Here, the background is
 dominated by the non-resonant spectator channels. Its value
 is almost similar for all considered values of $M_0$
 being much smaller than the total background at $M_0=1.52$~GeV.

 At $M_0=1.52$ GeV, the height of the peak
 of the coherent $\Lambda^*\Theta^+$ channel is about one third
 of the total background contribution. This ratio decreases for
 $M_0 = M_{\Lambda^*}\pm 70$ MeV. Thus, a summary plot of the
 total  $nK^+$ invariant mass  distribution
 for three fixed intervals of the  $pK^-$ invariant mass
 is shown in Fig.~\ref{fig:18}.
 \begin{figure}[h!]
 {\centering
 \includegraphics[width=.35\textwidth]{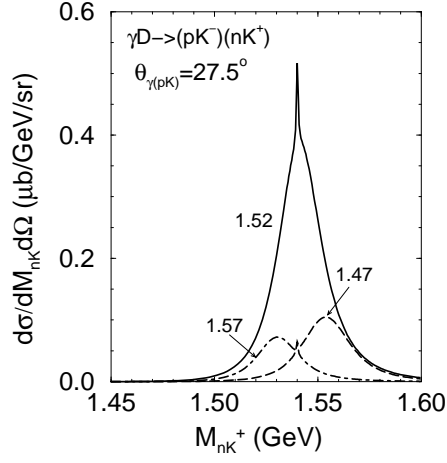}
 \caption{\label{fig:18}{\small
 A summary plot of the total  $nK^+$ invariant mass  distribution
 in the $\gamma D\to npK^+K^-$ reaction at
 three fixed intervals of the $pK^-$ invariant mass
 with $M_{0}=1.52$, $1.57$ and $1.47$ (GeV) at
 $\theta_{\gamma(pK^-)}=27.5^o$ and $E_\gamma=2$ GeV.}}}
 \end{figure}
 One can conclude that, since the width of the coherent
 photoproduction is much smaller than the effective width
 of the background, this contribution can be extracted
 experimentally under the condition of
 a high resolution measurement of the $nK^+$ invariant mass.

 In case of a energy resolution comparable to the width
 of the background peak one has to smear
 this peak.
 \begin{figure}[h!]
 {\centering
 \includegraphics[width=.4\textwidth]{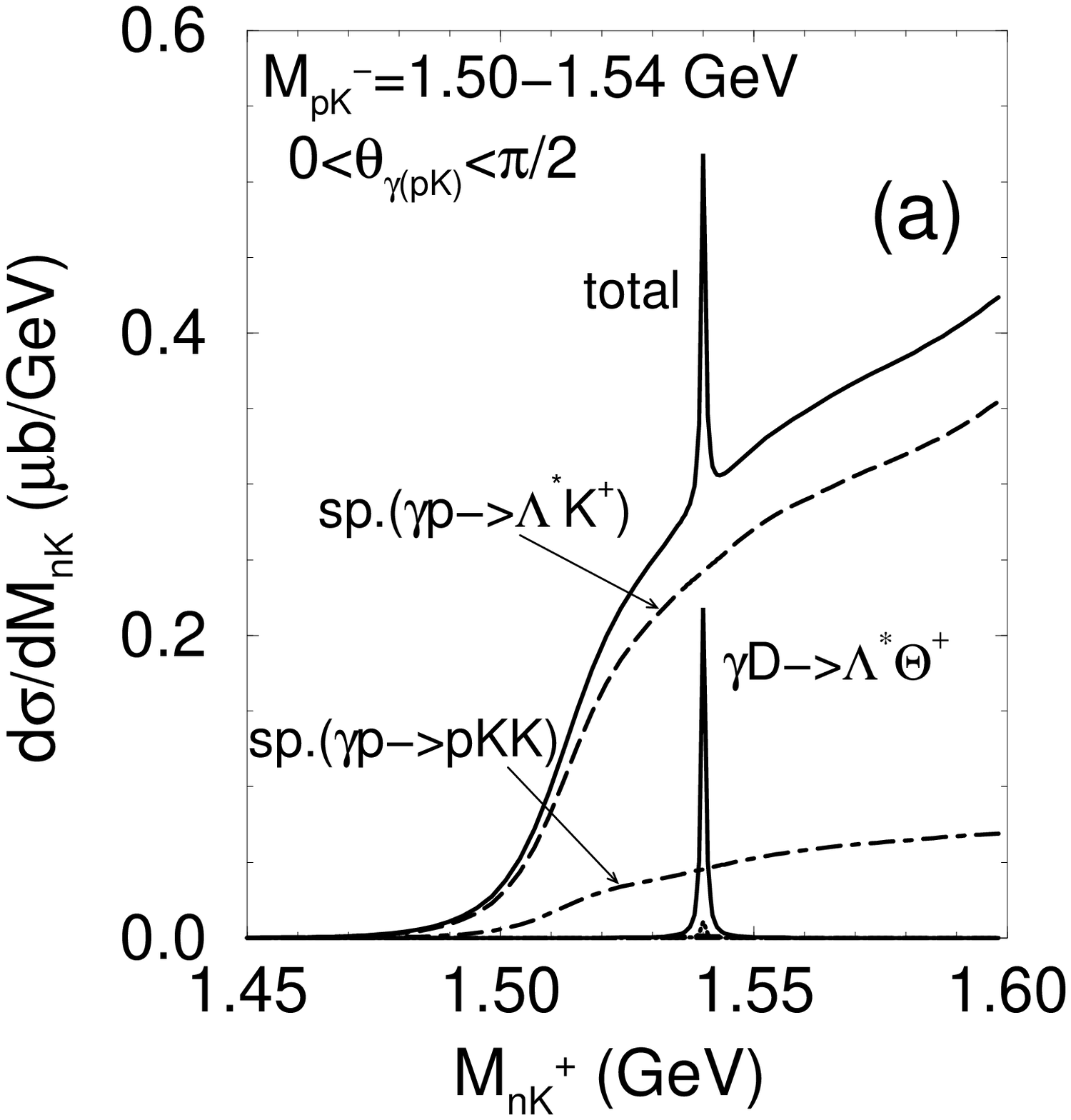}
 \includegraphics[width=.4\textwidth]{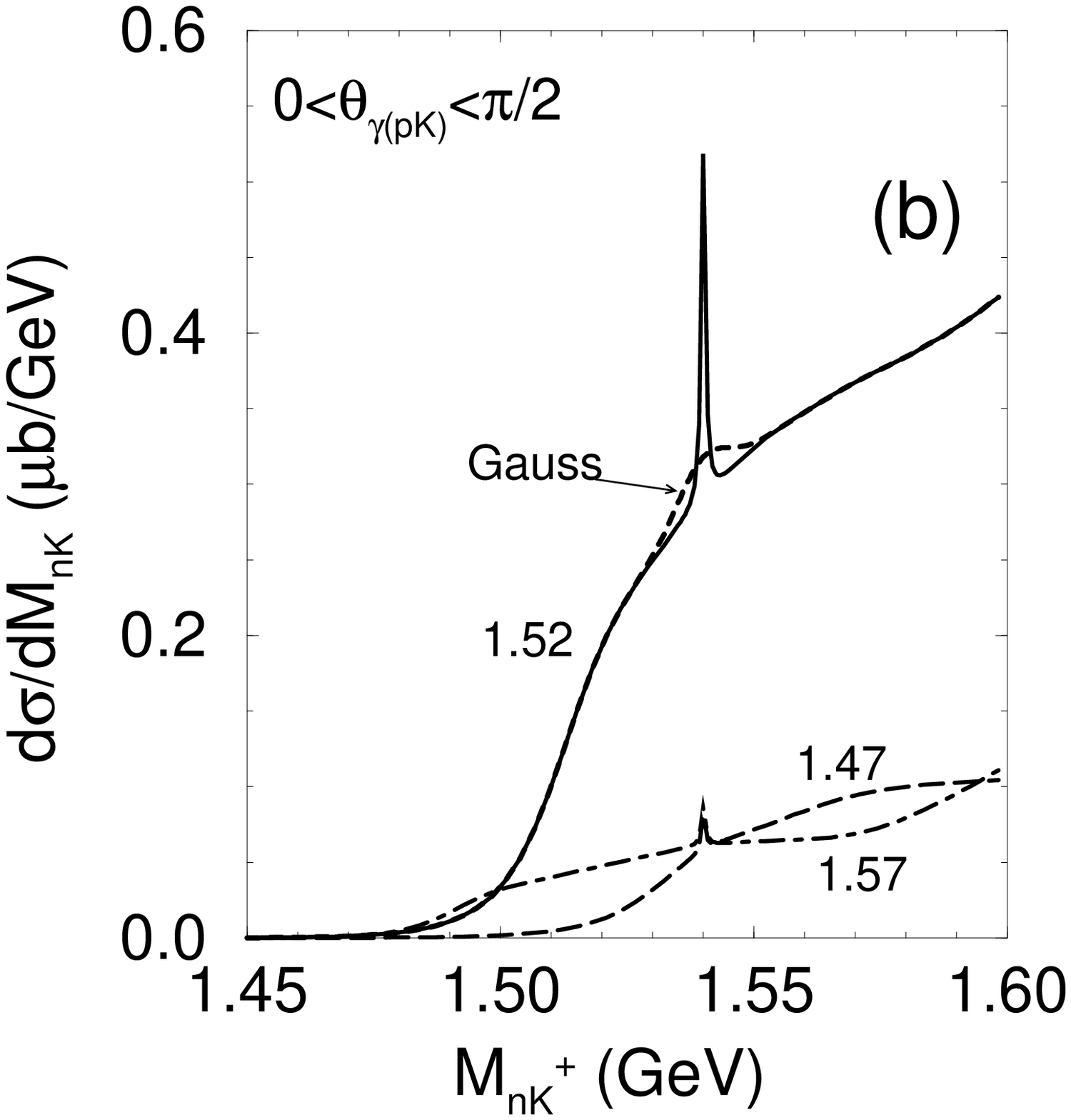}
 \caption{\label{fig:19}{\small
 (a) The  $nK^+$ invariant mass  distribution
 in the $\gamma D\to npK^+K^-$ reaction at
 $M_{0}=1.52$~GeV at
 the forward hemisphere of $pK^-$ pair photoproduction
 and $E_\gamma=2$ GeV.
 Notations are the same as in Fig.~\protect\ref{fig:17}.
 (b) A summary plot of the total  $nK^+$ invariant mass  distribution
 at three fixed intervals of the $pK^-$ invariant mass
 with $M_{0}=1.52$, $1.57$ and $1.47$ GeV.}}}
 \end{figure}
The simplest way to do it is integrating the $nK^+$ invariant mass
distribution  over $\Omega$ in the forward hemisphere of the
$pK^-$ pair photoproduction. The corresponding predictions for
$M_0=1.52$~GeV and a summary plot for three values of $M_0$ are
shown in Figs.~\ref{fig:19}a and b, respectively. One can see that
again at $M_0=M_{\Lambda^*}$ the background is dominated by the
resonance $\Lambda^*$ photoproduction where the neutron is a
spectator. But the shape of the background is quite different from
the previous case. Instead of the narrow peak one observes a
monotonous increase of the background contribution. This behavior
allows to extract the sharp $\Theta^+$ peak of the coherent
$\Lambda^*\Theta^+$ photoproduction. The peak becomes negligible
at $M_0=M_{\Lambda^*}\pm70$~MeV, as  shown in Fig.~\ref{fig:19}b.
Here one can also see the prediction for a Gaussian smearing of
the $\Theta^+$ peak with $\sigma=5$~MeV.

 Consider now the backward hemisphere of the $pK^-$ pair
 photoproduction in the reaction $\gamma D\to npK^+K^-$,
 say for  $\theta_{\gamma(pK^-)}\simeq152.5^o$.
 The corresponding
 invariant mass distributions at different $M_0$  are exhibited
 in Fig.~\ref{fig:20}.
 \begin{figure}[h!]
 {\centering
 \includegraphics[width=.3\textwidth]{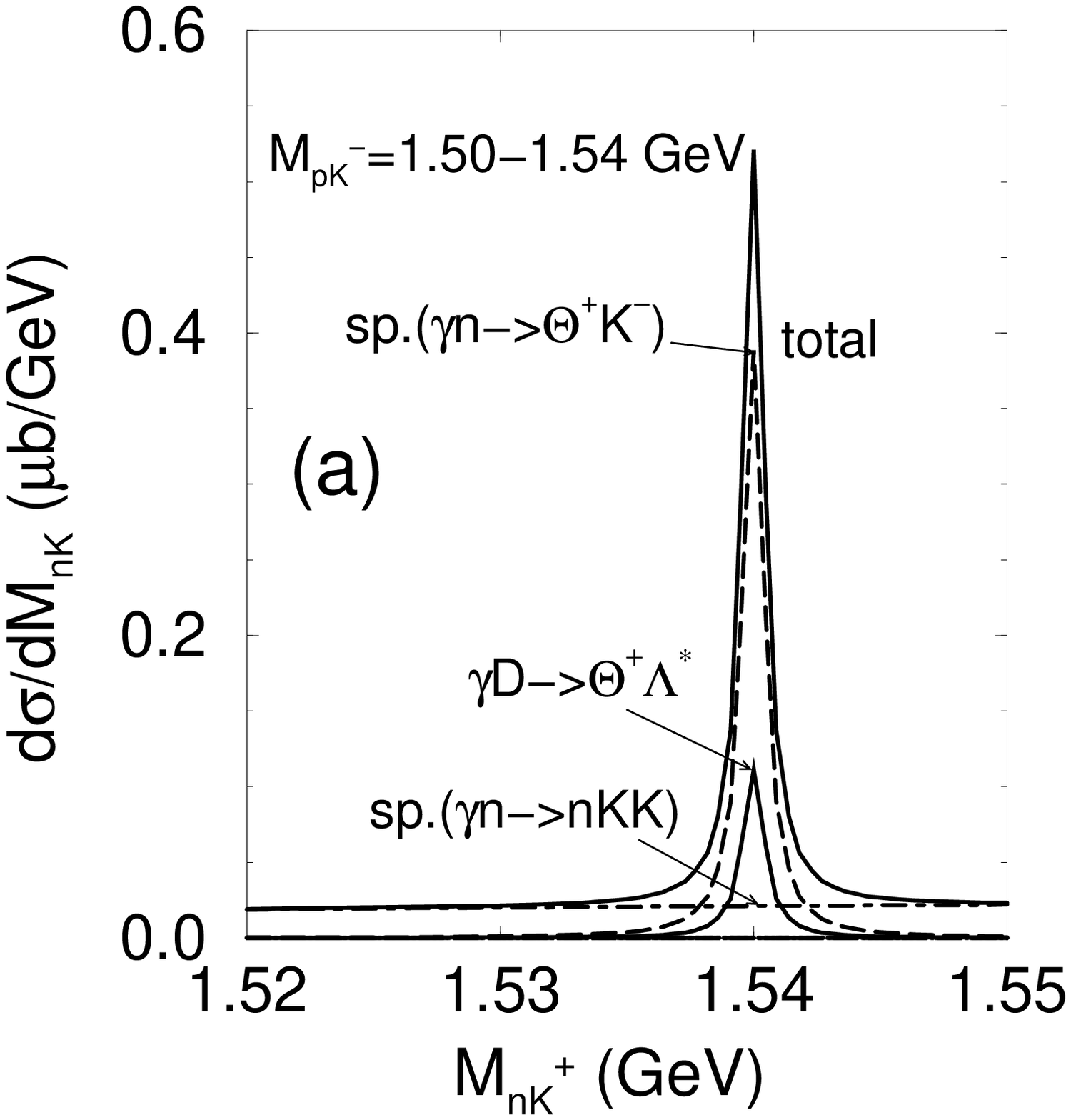}\qquad
 \includegraphics[width=.3\textwidth]{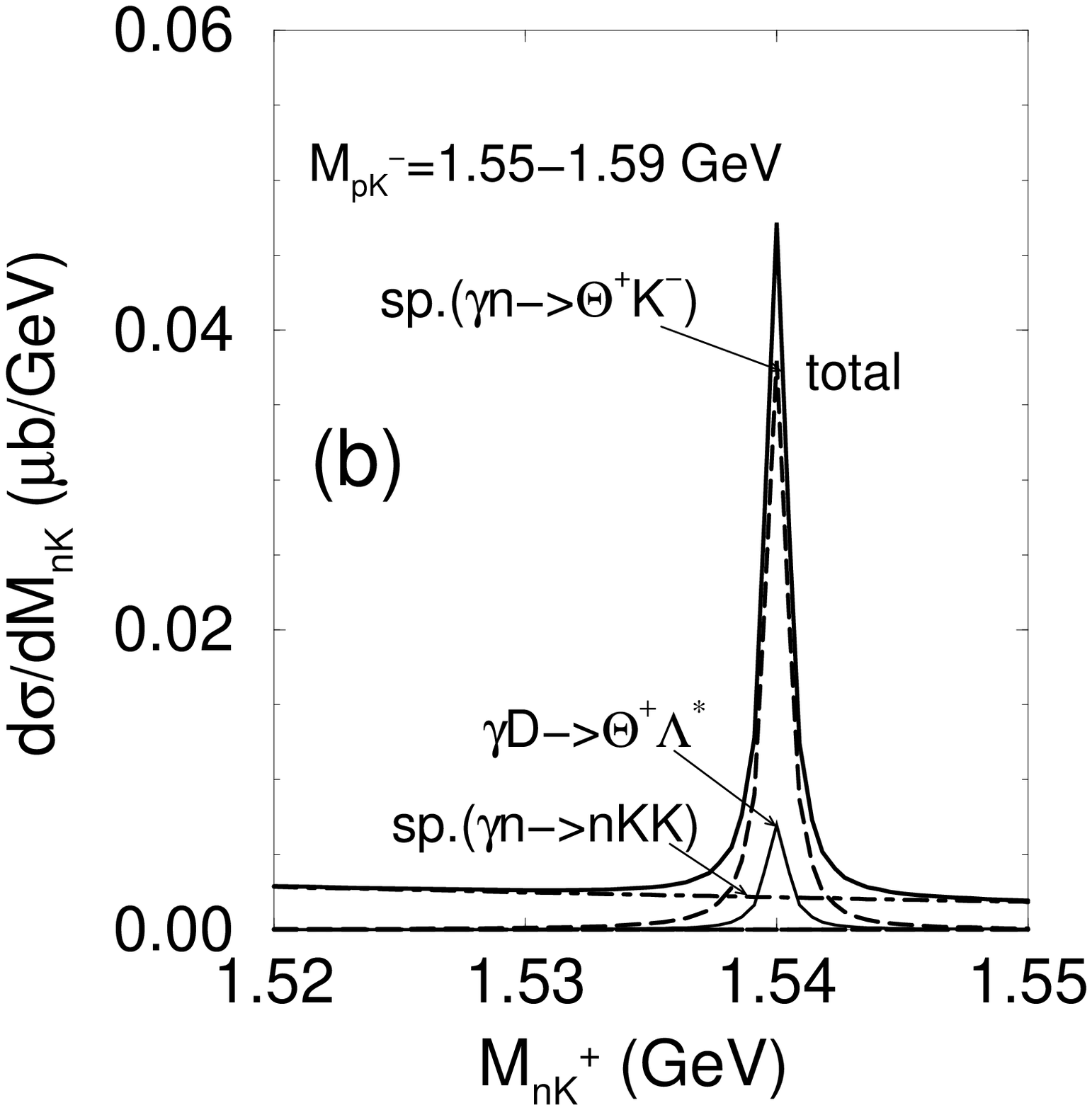}\qquad
 \includegraphics[width=.3\textwidth]{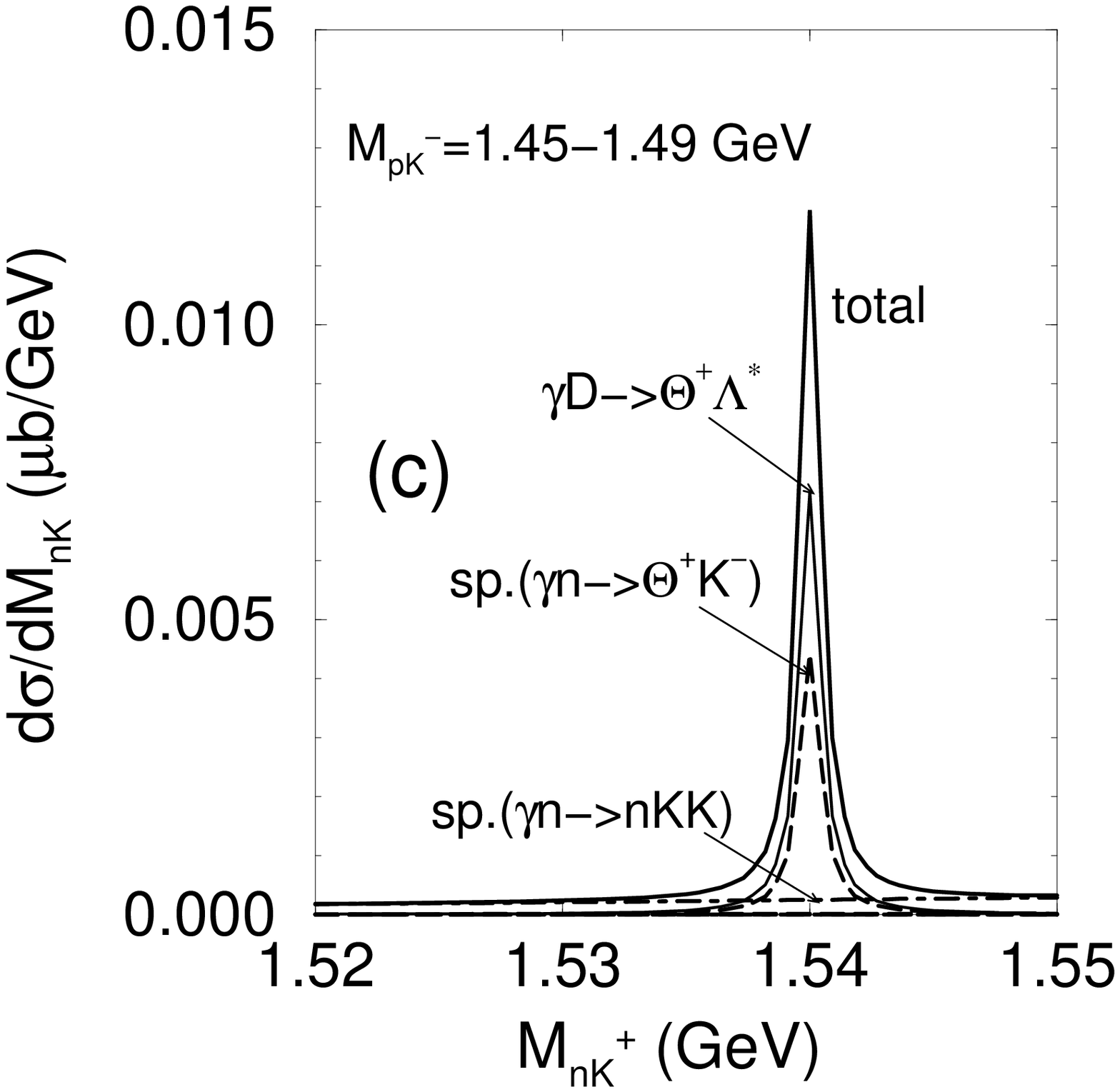}
 \caption{\label{fig:20}{\small
 The same as in Fig.~\protect\ref{fig:17} but
 for $\theta_{\gamma(pK^-)}=152.5^o$.
 Notations "$sp.(\gamma n\to \Theta^+K^+)$"
 and "$sp.(\gamma n\to nKK)$" corresponds to the processes
 depicted in Fig.~\protect\ref{fig:10}c and d, respectively.}}}
 \end{figure}
 Now, the dominant contribution to the background comes from the
 spectator resonant $\Theta^+$ photoproduction, depicted in
 Fig.~\ref{fig:10}c. The other channels are rather weak.
 At $M_0=1.52$~GeV the background contribution is enhanced by
 the distribution function $W_{pK}$ which at $M_{nK^+}\simeq1.54$
 GeV is much greater for $M_0\simeq M_{\Lambda^*}$
 (see Fig.~\ref{fig:14}b). The coherent contribution
 of the $\Lambda^*\Theta^+$ photoproduction is a factor of four
 smaller than the background contribution.

\begin{figure}[h!]
 {\centering
 \includegraphics[width=.35\textwidth]{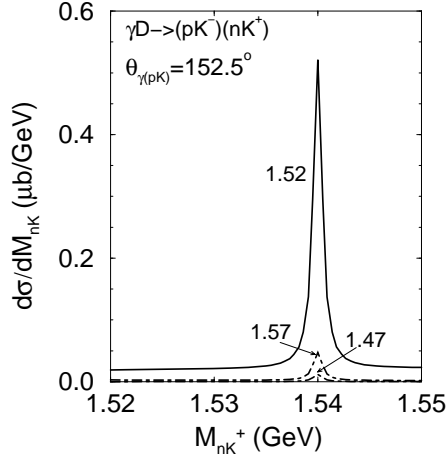}
 \caption{\label{fig:21}{\small
 The same as in  Fig.~\protect\ref{fig:18} but for $\theta_{\gamma(pK^-)}=152.5^o$.}}}
 \end{figure}

 The  summary plot of the
 total invariant mass  distribution of the $nK^+$
 for three fixed intervals of the  $pK^-$ invariant mass
 is displayed in Fig.~\ref{fig:21}. One can see a strong increase
 of the invariant mass distribution at $M_0=1.52$ GeV.
 But this increase is caused  mainly by the properties of the
 distribution function $W_{pK}$. Here, we have no striking
 qualitative effect of the coherent $\Lambda^*\Theta^+$
 photoproduction. Therefore,  studying the coherent $\Lambda\Theta^+$
 photoproduction seems to be difficult in this kinematical region.

\begin{figure}[h!]
 {\centering
 \includegraphics[width=.35\textwidth]{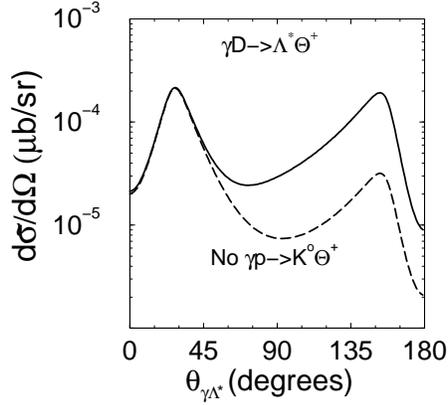}
 \caption{\label{fig:22}{\small
 The differential cross section
 of  $\Lambda^*\Theta^+$ photoproduction with (solid curve)
 and without (dashed curve) contribution of the $\gamma p\to \Theta^+\bar K^0$
 subprocess.}}}
 \end{figure}

 Now we would like to make three comments.
 First, since in the forward hemisphere of the $\Lambda^*$
 photoproduction the dominant contribution comes from the
 backward angles of the $K^+$ photoproduction in the elementary
 $\gamma p\to \Lambda^*\bar K^+$ subprocess,  our
 predictions are not sensitive to the choice of the solution
 for the coupling strength $\alpha_{\Lambda^*}$ discussed in
 Sec.~II (see Fig.~\ref{fig:6}~b).

 \noindent
 Second,
 in our analysis we have assumed that the $\Theta^+$
 photoproduction from the nucleon is dominated by the $t$-channel $K^*$
 exchange process. This assumption leads to a similarity of the
 $\Theta^+$ photoproduction from the neutron and proton. A
 violation
 of this similarity (or a suppression of the photoproduction from
 the proton, with keeping the cross section of the $\gamma n\to
 \Theta^+K^-$ on the same level) discussed
 recently~\cite{Hosaka0505,Vita} would result in a suppression of the
 process shown in Fig.~\ref{fig:7}d. As a consequence,
 the coherent cross section of $\Lambda^*\Theta^+$ photoproduction
 would be suppressed around the  second peak at backward angles
 of the $pK^-$ pair photoproduction, shown in Figs.~\ref{fig:9} and
 \ref{fig:15} leaving the first peak at forward angles photoproduction
 without change. The corresponding calculation of the differential cross section
 with and without contribution of the $\gamma p\to \Theta^+\bar K^0$
 subprocess is presented in Fig.~\ref{fig:22}. Since the coherent
 $\Lambda^*\Theta^+$ photoproduction is determined by the first
 peak, our main result shown in Fig.~\ref{fig:19}
 remains  unchanged.

 \noindent
 Third, the "bump-like" structure of the
 differential cross section of the coherent
 $\gamma D\to \Lambda^*\Theta^+$ reaction is caused mainly by the spectral
 functions $S$ in Eqs.~(\ref{gamma-D2}). Thus in Eq.~(\ref{gamma-D2E1}),
 the amplitude of the $\Theta^+\to nK^+$
 transition is a smooth function compared to the spectral function
 $S^{\Lambda^*}$ independently on the properties
 of $\Theta^+$. Therefore, our predictions remain
 to be valid for the $J^P=\frac{3}{2}^{\pm}$ of $\Theta^+$,
 considered in recent Ref.~\cite{Hosaka0505}.

{  When our prediction is to be compared with experiments, one
should pay attention, at least, the following two points. First,
an energy spread in the beam photon may change the shape of the
background, which is mainly determined by the quasi-free
$\Lambda^*$ production. However, our conclusion indicated by
Fig.~\ref{fig:19} is not changed qualitatively.
       Second, the shape of the background is sensitive to the
acceptance of the measurement. In particular, the effect of
coherent $\Lambda^* \Theta^+$ production may be significantly
suppressed when the detector does not have acceptance to detect $p
K^-$ pair in the forward angles. In contrast, the acceptance to
the forward $p K^-$ like one in the case of LEPS of
SPring-8~\cite{Nakano03} may make the effect more pronounced.}

\section{Summary}

In summary we analyzed the coherent  $\Lambda^*\Theta^+$
photoproduction in $\gamma D$ interaction with taking into account
different background processes. We found that the behavior and the
strength of the background processes depend strongly on the
kinematics where the momentum distribution in the deuteron plays a
key role. Thus, at fixed angle of the $pK^-$ photoproduction the
$nK^+$ invariant mass distribution of the background processes
looks like a narrow peak with maximum around the $\Theta^+$ mass.
This behaviour hampers the extraction of the coherent process at
finite invariant mass resolution. Most promising is an
experimental analysis of the distributions integrated over the
$pK^-$ production angles in the forward hemisphere of c.m.s. In
this case the background processes increase monotonously with
$M_{nK^+}$ in the vicinity of $M_{\Theta^+}$, which allows to
extract the coherent $\gamma D\to\Lambda^*\Theta^+$ channel even
with finite invariant mass resolution. We demonstrated that the
coherent $\Theta^+\Lambda(1520)$ photoproduction does not depend
on the $\Theta^+$ photoproduction amplitude, but rather it is
defined by the probabilities of the $\Lambda(1520)$
photoproduction and the $\Theta^+\to NK$ transition. Therefore,
this effect may be used as an independent method for studying the
mechanism of $\Theta^+$ production and $\Theta^+$ properties.

Our model estimates for the  $\gamma D$ reaction may be considered
as an example why the $\Theta^+$ peak is seen under certain
experimental conditions and why it does not appear above the
strong background in other ones.

Finally, we note that the predicted process of the coherent
$\Lambda^*\Theta^+$ photoproduction may be studied experimentally
at the electron and photon facilities at LEPS of SPring-8, JLab,
Crystal-Barrel of ELSA, and GRAAL of ESFR.

\acknowledgments

 We appreciate fruitful discussions with T.~Nakano who initiated this study,
 and we  thank
 H.~Ejiri, M.~Fujiwara, K.~Hicks and A.~Hosaka for useful comments and suggestions.
 One of authors (A.I.T.) thanks E.~Grosse for offering the hospitality at FZR.
 This work was supported by BMBF grant 06DR121, GSI-FE.

\appendix

\section{ Transition operators for the resonance amplitudes}

\subsection{\boldmath The $\Theta^+$ photoproduction amplitude}

We show here the explicit expressions for the transition operators
${\cal M}_\mu$ in Eq.~(\ref{res_ampl}) for a positive $\Theta^+$
parity  and the PS coupling scheme.

The specific parameters for the form factor in Eq.~(\ref{FF}) are
defined by
\begin{equation}
  F_s=F(M_N,s)~,\qquad  F_u=F(M_\Theta,u)~,
  \qquad   \text{and}\qquad
  F_t=F(M_{K^+},t)~.
\label{PVgn-plus-FF}
\end{equation}
In addition, we need the form factor combinations
\begin{equation}
  \widetilde{F}_{tu}=F_t+F_u -F_tF_u
  \qquad   \text{and}\qquad
  \widetilde{F}_{su}=F_s+F_u -F_sF_u
\end{equation}
to construct the contact terms $\mathcal{M}_\mu^c$ given below
that make the initial photoproduction amplitude gauge
invariant~\cite{hhgauge,DavWork}. The four-momenta in the
following equations are defined according to the arguments given
in the reaction equation
\begin{equation}
  \gamma(k)+N(p) \to \Theta^+(p_\Theta)+\bar{K}(\bar{q})~.
\end{equation}

\subsubsection{\boldmath $\gamma n\to \Theta^+ K^-$}
\begin{subequations}
\label{PSgn-plus}
\begin{align}
  {\cal M}^t_\mu&=i\frac{eg_{\Theta NK}(k_\mu-2\bar q_\mu)\gamma_5}{t-M^2_{K^+}}
 \,F_t~,\label{PSgn-plus-t}\\
 {\cal M}^s_\mu&=i{eg_{\Theta NK}}
  \gamma_5\frac{\fs p+\fs k +M_N}{s-M_N^2}
  \left(i\frac{\kappa_p}{2M_N}\sigma_{\mu\nu}k^\nu\right)\,F_s~,
 \label{PSgn-plus-s}\\
 {\cal M}^u_\mu&=i{eg_{\Theta NK}}
  \left(\gamma_\mu + i\frac{\kappa_\Theta}{2M_\Theta}\sigma_{\mu\nu}k^\nu\right)
  \frac{\fs p_{\Theta}-\fs k + M_\Theta}{u - M_\Theta^2} \gamma_5\,F_u~,
 \label{PSgn-plus-u}
 \displaybreak[1]
\\
 {\cal M}^c_\mu&=i{eg_{\Theta NK}\gamma_5}
 \left[\frac{(k-2\bar q)_\mu}{t-M^2_{K^+}}
 \,(\widetilde{F}_{tu}-F_t)
 +\frac{(2{p_\Theta}-k)_\mu}{u - M_\Theta^2}\,(\widetilde{F}_{tu}-F_u)\right]~.
 \label{PSgn-plus-c}
\end{align}
\end{subequations}
 The transition operator of $t$-channel $K^*$ exchange amplitude is given by
\begin{eqnarray}
 {\cal M}^t_\mu(K^*)=\frac{eg_{\gamma KK^*}g_{\Theta NK^*}}{M_{K^*}}\,
  \frac{\varepsilon_{\mu\nu\alpha\beta}k^\alpha \bar q^\beta}{t-M^2_{K^*}}
  \left[\gamma^\nu
  -i\frac{\sigma^{\nu\lambda}(p-p_\Theta)_\lambda}{M_\Theta+M_N} \kappa^*\right]\,F(M_{K^*},t)~.
 \label{gn-plusK*}
\end{eqnarray}

\subsubsection{\boldmath $\gamma p\to \Theta^+ \bar{K}^0$}
\begin{subequations}
\label{PVgp-plus}
\begin{align}
{\cal M}^s_\mu&=i\frac{eg_{\Theta NK}}{M_{\Theta}+M_N}\,
  \gamma_5\fs{\bar{q}}\frac{\fs p+\fs k +M_N}{s-M_N^2}
  \left(\gamma_\mu +i\frac{\kappa_p}{2M_N}\sigma_{\mu\nu}k^\nu\right)F_s~,
 \label{PVgp-plus-s}\\
 {\cal M}^u_\mu&=i\frac{eg_{\Theta NK}}{M_{\Theta}+M_N}\,
  \left(\gamma_\mu + i\frac{\kappa_\Theta}{2M_\Theta}\sigma_{\mu\nu}k^\nu\right)
  \frac{\fs p_{\Theta}-\fs k + M_\Theta}{u - M_\Theta^2} \gamma_5\fs{\bar{q}}\,F_u~,
 \label{PVgp-plus-u}\\
 {\cal M}^c_\mu&=
 i\frac{e g_{\Theta NK} }{M_\Theta+M_N} \gamma_5 \fs{\bar{q}}\left[
  \frac{(2{p} +k)_\mu} {s - M_N}\,(\widetilde{F}_{su}-F_s)
  + \frac{(2{p_\Theta} - k)_\mu } {u -
  M_\Theta^2}\,(\widetilde{F}_{su}-F_u)
  \right] ~.
 \label{PVgp-plus-c}
\end{align}
\end{subequations}

\subsection{\boldmath $\Lambda^*$ photoproduction amplitude}

  We show here the explicit expressions for the transition operators
  ${\cal M}_{\sigma\mu}$ in Eq.~(\ref{res_amplL}) for the reactions
  $\gamma p\to \Lambda^* K^+$ and   $\gamma n\to \Lambda^* K^0$.

\subsubsection{\boldmath $\gamma p\to \Lambda^* K^+$}
\begin{subequations}
\label{MtransL-p}
\begin{align}
  {\cal M}^t_{\sigma\mu}&=
  i\frac{eg_{\Lambda^* NK}{M_\Lambda^*}(2q_\mu-k_\mu)(k_\sigma-q_\sigma)\gamma_5}{t-M^2_{K^+}}
 \,F_t~,\label{L-p-t}\\
 {\cal M}^s_{\sigma\mu}&=-i
 \frac{eg_{\Lambda^* NK}}{M_\Lambda^*}\,q^\sigma
  \gamma_5\frac{\fs p+\fs k +M_N}{s-M_N^2}
  \left(\gamma_\mu +i\frac{\kappa_p}{2M_N}\sigma_{\mu\nu}k^\nu\right)F_s~,
 \label{L-p-s}\\
 {\cal M}^c_{\sigma\mu}&= i\frac{eg_{\Lambda^* NK}}{M_\Lambda^*}\,  \gamma_5
 \left[\frac{(2q-k)_\mu(k-q)_\sigma}{t-M^2_{K^+}}
 \,(\widetilde{F}_{ts}-F_t)
 - \frac{(2p + k)_\mu}{s -
 M_N}\,(\widetilde{F}_{ts}-F_s)\right.\nonumber\\
 &\left. \qquad\qquad \qquad\qquad \qquad\qquad \qquad\qquad
  +g_{\sigma\mu}\,\widetilde{F}_{ts}  \right]~.
 \label{L-p-c}
\end{align}
\end{subequations}
 The corresponding form factors are defined by
\begin{equation}
  F_s=F(M_N,s)~,\qquad
  F_t=F(M_{K^+},t)~,\qquad\text{and}\qquad
   \widetilde{F}_{ts}=F_t+F_s -F_tF_s~.
\label{L-FF}
\end{equation}
 The transition operator of $t$-channel $K^*$ meson exchange amplitude is given by
\begin{eqnarray}
 {\cal M}^t_{\sigma\mu}(K^*)=\frac{eg_{\gamma KK^*}g_{\Lambda^* NK^*}}{M_{K^*}M_{\Lambda^*}}\,
  \frac{\varepsilon_{\nu\mu\alpha\beta}\,k^\nu q^\alpha}{t-M^2_{K^*}}
  \left[ q_\sigma' \gamma_\sigma - \fs{q}'g_{\sigma\beta}\right]\,F(M_{K^*},t)
 \label{gp-plusK*}
\end{eqnarray}
with $q'=p_{\Lambda^*} - p$.

\subsubsection{\boldmath $\gamma n\to \Lambda^*{K}^0$}
\begin{eqnarray}
 {\cal M}^s_{\sigma\mu}=-i
 \frac{eg_{\Theta NK}}{M_\Lambda^*}\,q^\sigma
 \gamma_5\frac{\fs p+\fs k +M_N}{s-M_N^2}
 \left( i\frac{\kappa_p}{2M_N}\sigma_{\mu\nu}k^\nu\right)F_s~.
 \label{L-n-s}
\end{eqnarray}
The $t$-channel $K^*$-exchange operator is defined by
Eq.~(\ref{gp-plusK*}) with appropriate coupling constant
$g_{\gamma KK^*}$.


\begin{thebibliography}{99} 
 \bibitem{Nakano03}
  T. Nakano \textit{et al.} [LEPS Collaboration], Phys.\ Rev.\ Lett.\
  \textbf{91}, 012002 (2003).
  \bibitem{OtherPenta}
  V.\,V.~Barmin \textit{et al.} [DIANA Collaboration], Phys.\ Atom.\ Nucl.\
  \textbf{66}, 1715 (2003);\\
  S. Stepanyan \textit{et al.} [CLAS Collaboration], Phys.\ Rev.\ Lett.\ \textbf{91},
  252001 (2003);\\
  V.~Kubarovsky, S.~Stepanyan \textit{et al.} [CLAS Collaboration],
  Phys.\ Rev.\ Lett.\ \textbf{92},
  032001 (2004);\\
  J.~Barth \textit{et al.} [SAPHIR Collaboration], Phys.\ Lett.\ B \textbf{572},
  127 (2003);\\
   A.\,E.~Asratyan, A.\,G.~Dolgolenko, and M.\,A.~Kubantsev,
    Phys.\ Atom.\ Nucl.\ \textbf{67}, 682 (2004), Yad.\ Fiz.\ \textbf{67}, 704
    (2004);\\
   L.~Camiller \textit{et al.} [SAPHIR Collaboration], Phys.\ Lett.\ B \textbf{572},
  127 (2003).

\bibitem{Hicks}
 K. Hicks,
 arXiv:hep-ex/0504027,
 Submitted to Prog.\ Part.\ Nucl.\ Phys.

\bibitem{Kabana05}
S.~Kabana, J.\ Phys.\ G {\bf 31}, S1155 (2005).

\bibitem{NakanoP04}
T.~Nakano, in Proc.\ of Int.\ Workshop PENTAQUARK04, Spring-8
Japan, 20 - 23 July 2004. Ed. by A.~Hosaka and T.~Hotta. World
Scientific 2005; http://www.rcnp.osaka-u.ac.jp/penta04/.

\bibitem{TedeschiP04}
 D.J. Tedeschi,
in Proc.\ of Int.\ Workshop PENTAQUARK04, Spring-8 Japan, 20 - 23
July 2004. Ed. by A.~Hosaka and T.~Hotta. World Scientific 2005.

\bibitem{Hosaka0503}
S.~I.~Nam, A.~Hosaka,  and H.~C.~Kim, arXiv:hep-ph/0503149.

\bibitem{TEHN04}
 A.\,I.~Titov, H.~Ejiri, H.~Haberzettl, and K.~Nakayama,
 Phys.\ Rev.\ C \textbf{71}, 035203 (2005).

\bibitem{Hosaka0505}
S.~I.~Nam, A.~Hosaka, and H.~C.~Kim, arXiv:hep-ph/0505134.

\bibitem{SmallWidth}
 R.\,A.~Arndt, I.\,I.~Strakovsky, and R.\,L.~Workman, Phys.\ Rev.\
 C \textbf{68}, 042201(R) (2003),  Erratum-\textit{ibid.\/} \textbf{69},
 019901(E) (2004);\\
 J.~Haidenbauer and G.~Krein, Phys.\ Rev.\ C \textbf{68},
 052201(R) (2003);
 A.~Sibirtsev, J. Haidenbauer, S. Krewald, and U.-G. Meissner,
 Phys.\ Lett.\ B \textbf{599}, 230 (2004);
 A.~Sibirtsev, J.~Haidenbauer, S.~Krewald,
 and U.-G. Meissner, Eur.\ Phys.\ J.\ A {\bf 23}, 491 (2005);
 A.~Casher and S.~Nussinov,
 Phys.\ Lett.\ B \textbf{578}, 124 (2004)

\bibitem{PDG}
S.~Eidelman \textit{et al.} [Particle Data Group Collaboration],
Phys. Lett.\ B \textbf{592}, 1 (2004).
\bibitem{NT03}
 K.~Nakayama and K.~Tsushima, Phys.\ Lett.\ B \textbf{583}, 269
 (2004).

\bibitem{Zhao03}
 Q.~Zhao, Phys.\ Rev.\ D \textbf{69}, 053009 (2004),
  Erratum-\textit{ibid.\/} D \textbf{70}, 039901 (2004).

\bibitem{ZhaoKhal04}
 Q.~Zhao and J.\,S.~Al-Khalili,
 Phys.\ Lett.\ B \textbf{585}, 91 (2004),
 Erratum-\textit{ibid.\/} B \textbf{596}, 317, (2004).

\bibitem{Hosaka03}
 S.\,I.~Nam, A.~Hosaka, and H.\,C.~Kim, Phys.\ Lett.\ B \textbf{579}, 43 (2004).

\bibitem{OKL031}
 Y.~Oh, H.~Kim, and S.\,H.~Lee, Phys.\ Rev.\ D {\bf69}, 014009
 (2004).

\bibitem{LiuKo031}
 W.~Liu and C.\,M.~Ko, Phys.\ Rev.\ C \textbf{68}, 045203
 (2003);\\
 W.~Liu and C.\,M.~Ko, Nucl.\ Phys.\ A \textbf{741}, 215 (2004);\\
 W.~Liu, C.\,M.~Ko, and V.~Kubarovsky,
 Phys.\ Rev.\ C \textbf{69}, 025202 (2004).

\bibitem{Oh-2}
 Y.~Oh, H.\,C.~Kim, and S.\, H.~Lee,
 Nucl.\ Phys.\ A \textbf{745}, 129 (2004).

 \bibitem{CloseZhao}
 F.\,E.~Close and Qiang Zhao,
 Phys.\ Lett.\ B \textbf{590}, 176 (2004).

 \bibitem{Roberts04}
 W.~Roberts, Phys.\ Rev.\ C {\bf 70}, 065201 (2004).

\bibitem{Mart:2004at}
T.~Mart, Phys.\ Rev.\ C {\bf 71}, 022202(R) (2005).

\bibitem{Oh:2004wp}
Y.~Oh, K.~Nakayama and T.~S.~Lee, arXiv:hep-ph/0412363.

\bibitem{Vita}
R.~De~Vita \textit{et al.} [CLAS Collaboration], talk given at APS
meeting April (2005).

\bibitem{gammaD}
 V. Guzey,
 Phys.\ Rev.\ C\ \textbf{69}, 065203 (2004).

\bibitem{hhgauge}
H.~Haberzettl, Phys.\ Rev.\  C~\textbf{56}, 2041 (1997);\\
H.~Haberzettl, C.~Bennhold, T.~Mart, and T.~Feuster, Phys.\ Rev.\
C~\textbf{58}, R40 (1998).

\bibitem{DavWork}
R.\,M.~Davidson and R.~Workman, Phys.\ Rev.\ C {\bf63}, 025210
(2001).

\bibitem{MagMom}
Y.-R.~Liu, P.-Z.~Huang, W.-Z.~Deng, X.-L.~Chen, and Shi-Lin~Zhu,
Phys.\ Rev.\ C \textbf{69}, 035205 (2004).

\bibitem{QMKK*}
 C.\,E.~Carlson, C.\,D.~Carone, H.\,J.~Kwee, and V.~Nazaryan,
 Phys.\ Rev.\ D \textbf{70}, 037501 (2004);
 F.\,E~Close and J.\,J~Dudek,
 Phys.\ Lett.\ B \textbf{586}, 75 (2004).

 \bibitem{LambdaSigma}
 S.~Janssen, J.~Ryckebusch, D.~Debruyne, and T.~Van Cauteren,
 Phys.\ Rev.\ C \textbf{65}, 015201 (2002);
 \textit{ibid.\/} \textbf{66}, 035202 (2002).

 \bibitem{TitovLee}
 A.I.~Titov and T.-S.H.~Lee,
 Phys.\ Rev.\ C \textbf{66}, 015204 (2002).

\bibitem{3-2R}
 M.~Benmerrouche, N.\,C. Mukhopadhyay,  and J.\,F.~Zhang,
 Phys.\ Rev.\ D \textbf{51}, 3237 (1995).

\bibitem{Barber1980}
D.\,P.~Barber {\it et al.},
Z.\ Phys.\ C {\bf 7}, 17 (1980).

\bibitem{Collins}
P.D.B. Collins, "An Introduction to Regge Theory and High Energy
Physics", Cambridge University Press, 1977.

\bibitem{Paris}
 M.~Lacombe, B.~Loiseau, R.~Vinh Mau, J.~Cote, P.~Pires,
 and R.~de~Tourreil, Phys. Lett. B {\bf 101}, 139 (1981);\\
 M.~Lacombe, B.~Loiseau, J.M.~Richard, R.~Vinh Mau, J.~Cote,P.~Pires, and R.~de~Tourreil,
Phys.\ Rev.\ C {\bf 21}, 861 (1980).

\bibitem{exp-gp-KK}
Aachen-Berlin-Bonn-Hamburg-Heidelberg-M\"unchen Collaboration,
Phys.\ Rev.\ {\bf 188}, 2060 (1969).


\end{thebibliography}
\end{document}